\documentclass[final,authoryear,3p,times,twocolumn]{elsarticle}
 \usepackage{graphicx}
\usepackage{amssymb, amsmath}
 \usepackage{amsthm}

\newcommand{\mfrac}[2]{\frac{\displaystyle #1}{\displaystyle #2}}

\journal{Journal of Theoretical Biology}

\usepackage{fancyhdr}
\begin{document}

\begin{frontmatter}


%%\fntext[label2]{\bf Published in Journal of theoretical biology 263 (2010) 120-133}

\title{Vertical distribution and composition of phytoplankton under the influence of an upper mixed layer}


\author[icbm]{Alexei B. Ryabov \corref{cor1}}
\ead{a.ryabov@icbm.de}

\author[CS]{Lars Rudolf}
\ead{rudolf@mpipks-dresden.mpg.de}

\author[icbm]{Bernd Blasius}
\ead{blasius@icbm.de}

\cortext[cor1]{Corresponding author}

\address[icbm]{ICBM, University of Oldenburg, 26111 Oldenburg, Germany}
\address[CS]{Max Planck Institute for the Physics of Complex Systems, D-01187 Dresden, Germany}
\address[w2]{ {\bf Journal of Theoretical Biology 263 (2010) 120-133}}

\begin{abstract}
The vertical distribution of phytoplankton is of fundamental importance for the dynamics and structure of aquatic communities.
Here, using an advection-reaction-diffusion model, we investigate the distribution and competition of phytoplankton species in a water column, in which inverse resource gradients of light and a nutrient can limit growth of the biomass.  This problem poses a challenge for ecologists, as the location of a production layer is not fixed, but rather 
depends on many internal parameters and environmental factors.
In particular, we study the influence of an upper mixed layer (UML) in this system and show that it leads to a variety of dynamic effects:
(i) Our model predicts alternative density profiles with a maximum of biomass either within or below the UML, thereby the system may be bistable
or the relaxation from an unstable state may require a long-lasting transition.  
(ii) Reduced mixing in the deep layer can induce oscillations of the biomass; we show that a UML can sustain these oscillations even if the diffusivity is less than the critical mixing for a sinking phytoplankton population.
%, thus providing a novel mechanism to resolve the drift paradox.
%
(iii) A UML can strongly modify the outcome of competition between different phytoplankton species, yielding bistability both in the spatial distribution and in the species composition.
(iv) A light limited species can obtain a competitive advantage if the diffusivity in the deep layers is reduced below a critical value. This yields a subtle competitive exclusion effect, where 
the oscillatory states in the deep layers are displaced by steady solutions in the UML.
Finally, we present a novel graphical approach for deducing the competition outcome and for the
analysis of the role of a UML in aquatic systems.
\end{abstract}


\end{frontmatter}

\section{Introduction}
%\linespread{1}
The survival and competition of species in an heterogeneous environment has fascinated ecologists for long times (see e.g. Holmes et al. 1994; Tilman and Kareiva 1997; Huisman et al. 1999; Neuhauser 2001).
In many systems the spatial diversity of natural populations originates mainly from some underlying abiotic heterogeneity of the environment. If growth conditions vary between different locations this spatial variation should be reflected in the density distribution of natural populations.
After the seminal papers by Skellam (1951) and by Kierstead and Slobodkin (1953), the dynamics of such systems have often been analyzed in terms of favorable and unfavorable patches (see e.g. Okubo and Levin 2001; Cantrell and Cosner 2001; Birch et al. 2007). This approach assumes that space can be divided into regions of positive and negative net growth, between which the organisms are transported by diffusion and advection.
These suggestions, although realistic in many situations, do not hold in some resource-consumer systems, in which the size, the form and even the location of the species' favorable patch may vary, reflecting external and internal perturbations.

An important example is provided by the dynamics of phytoplankton in an incompletely mixed water column (see e.g. Jamart et al. 1977; Klausmeier and Litchman 2001; Beckmann and Hense 2007).  
In many regions of the world's ocean the form of vertical phytoplankton profiles is determined mainly by two factors.
First, the reduction of the light intensity with depth makes deep layers unfavorable for photosynthetic phytoplankton species. Second, an opposing gradient of nutrients can often maintain a positive net production only in deep subsurface layers.
%Thereby, light limitation leads to maxima of phytoplankton close to the surface, whereas lower nutrient concentrations favor phytoplankton build-up in deeper layers (Venrick 1973, 1993). 
These deep states, known as deep chlorophyll maxima (DCM), constitute one of the most striking characteristics of nutrient poor waters in ocean ecosystems and freshwater lakes  (Steele and Yentsch 1960; Anderson 1969; Abbott et al. 1984; Cullen 1982; Tittel et al. 2003).
The production layer in such systems is highly variable and dynamic.
First it is species specific: less nutrient limited species will have a positive production rate close to the surface, whereas smaller light requirements will result in subsurface productivity (see e.g. Venrick 1993). 
Second, as each species shades light and consumes nutrients it acts as an ecosystem-engineer, modifying its abiotic environment (Jones et al. 1994).
As a consequence, the position of the favorable layer for a phytoplankton species depends on the current abundance and distribution of its own biomass, as well as that of all other species.

%and changing the resource availabilities for other species.  

%It has been shown that the mobility of the favorable layer can lead to new types of dynamics (Yoshiyama and Nakajima 2002; Huisman et al. 2006) and new aspects for the species competition (Britton and Timm 1993). Nevertheless this problem still has not received much attention in the literature.This is astonishing, as phytoplankton is the primary producer in almost all aquatic foodwebs with a major influence on nearly all freshwater and marine ecosystems. Recently the connection with global climate change has underlined the importance for gaining a better understanding of the spatial distribution and composition of phytoplankton (e.g., Raven and Falkowski 1999; Klausmeier and Litchman 2001; Tittel et al. 2003; Huisman et al. 2006).

Theoretical models have been a useful approach to describe 
%and understand the dynamics of 
nutrient limited phytoplankton growth in constant and seasonally driven environments (Huppert et al. 2002, 2005). The dynamics, competition and structuring within a vertical water column have been investigated
in a series of modeling investigations (Radach and Maier-Reimer 1975; Shigesada and Okubo 1981; Varela et al. 1992; Huisman and Weissing 1995; Huisman et al. 1999; Klausmeier and Litchman 2001; Diehl 2002; Hodges and Rudnick 2004; Huisman et al. 2006; Beckmann and Hense 2007). 
It was shown that a given set of parameters may lead to either a surface or a deep chlorophyll maximum, whereby the location of the maximum is entirely determined by the environmental conditions. 
%If sinking of phytoplankton is taken into account, the population can only survive if diffusivity is larger than a critical value (Riley 1949; Shigesada and Okubo 1981; Hershey et al. 1993; Huisman et al. 2002).
These model results are in agreement with many field studies (e.g., Aristegui et al.  2003;  Matondkar et al. 2005; Weston et al. 2005). However, in a few recent investigations either surface or deep maxima were observed under almost the same conditions (Venrick 1993; Holm-Hansen and Hewes 2004).
These observations suggest the existence of bistability in the spatial phytoplankton configuration, however they cannot easily be reproduced in current models and may indicate that some important mechanisms, which are contributing to the spatial organization of aquatic communities, are still not well understood.

One major ingredient determining whether a species can establish close to the surface is the presence of an upper mixed layer (UML). 
UMLs commonly occur in oceans and lakes due to mechanical perturbation of the surface waters (e.g., by wind, waves, storms) and are characterized by strong turbulent mixing
up to a depth of about 30~m to 200~m or more (Deuser 1987; Venrick 1993; Law et al. 2000). 
%
%, in contrast to the rather slow diffusion transport in the deeper layers. 
%In the ocean the depths of a UML can vary from 30~m to 200~m or more (Deuser 1987; Venrick 1993; Law et al. 2000). 
%, but it may be much smaller in lakes and shallow waters. 
%A UML 
%%is one of the major components influencing the aquatic environment close to the surface. It 
%provides almost uniform distributions of biomass and resources (with the exception of light). Sinking %phytoplankton species can benefit from a UML, as they require a sufficient level of diffusivity to %prevent the wash-out from a favorable area. 
%%In contrast, a deep UML may result in the extinction of light limited species (Huisman et al. 1999b). 
%%Nevertheless, previous theoretical approaches showed that a UML has practically no effects on the %dynamics of phytoplankton if phytoplankton self-shading (Fennel and Boss 2003; Hodges and Rudnick 2004) %or the heterogeneity of the nutrient distribution (Huisman et al. 1999) are negligible.
%
%
To our knowledge the first study  to demonstrate the strong influence of a UML on the  spatial configuration of phytoplankton was performed  by Yoshiyama and Nakajima (2002, 2006), who considered a single species model where the water column was divided into an infinitely mixed UML and poorly mixed lower layers, with a small diffusivity across the separating boundary layer (thermocline). The model outcome were vertical phytoplankton patterns with a sharp edge at the thermocline, with the possibility of bistability in the phytoplankton distribution, characterized by phase transitions between biomass maxima in the deep and upper layers.
However, these articles did not analyze such important questions as the influence of finite mixing in the upper layer, the competition of species occupying different layers, or the dynamics of the system when deep phytoplankton maxima are not stationary, and thus many important questions about the role of the UML for structuring phytoplankton configurations still remain open.

In this study, we 
%investigate the influence of a UML on the vertical density profile and species composition of phytoplankton. We 
consider a mathematical model for the growth of a nutrient and light limited phytoplankton community in a vertical incompletely mixed water column.
%In contrast to Yoshiyama and Nakajima (2002, 2006) 
%In our model 
A UML is introduced by smoothly varying the strength of diffusivity with vertical depth, so that discontinuities are avoided.
%We observe a constructive interplay between a mobile production layer and the localized region with large diffusivity, yielding a variety of dynamical and ecological effects.
%Most notably, 
We find that a UML ameliorates the growth conditions close to the surface and 
%. Even though the spatial extension of the UML may be quite small compared to the rest of the water body, this 
can have drastic effects because organisms close to the surface occlude light and prevent growth in all deeper layers.
Thus the presence or absence of a UML turns out to be a major factor controlling the vertical distribution and competition outcome in the whole water column.

In particular, the presence of a UML yields the following key model outcomes:
First, the spatial density profile can become bistable with vertical maxima either close to the surface or in deep layers.  Close to the bistability range dynamics are characterized by time scale separation and the relaxation from an unstable state may require a long-lasting transition. 
Second, in the presence of a UML the persistence threshold of sinking phytoplankton disappears.
Thus, our model provides a new mechanism for the persistence of a population in a flow
%, so-called drift paradox 
(Hershey et al. 1993; Huisman et al. 2002; Ryabov and Blasius 2008).
%a population can persit to resolve the drift paradox, i.e., the persistence of a population in the upper reaches of steams, even though the organisms drift downwards .
Third, a UML can strongly modify the competition outcome between different phytoplankton species by providing a vertical niche for species which are better adapted to the conditions close to the surface. 
As a result, in the two species model, we observe bistability both in the spatial distribution and in the species composition.
Fourth, we show that a light limited species can obtain a competitive advantage if the diffusivity in the deep layers is reduced below a critical value. 
In this case we identify a dynamic competitive exclusion effect, where the species composition is determined by the dynamic state such that oscillatory states in the deep layers are displaced by steady solutions in the UML.
%All these effects are analyzed by numerical and analytical means. We classify the typical model outcomes of the single-species and the two-species model and use heavy numerics to indicate the corresponding dynamic regimes in several parameter planes.
Finally, we present a graphical approach for analysing the
%which in many cases helps to gain insight into 
mechanism of resource competition in a spatially extended system.

%understand mechanisms of competition in spatially extended systems. %the competition outcome from the equilibrium pattern of each species alone.

\section{Model \label{sec:model}}

Our model describes the dynamics of $n$ phytoplankton species, competing for resources in a vertical water column of depth $Z_B$ (in the following we consider only the cases $n=1$ and $n=2$).
We study a bottom-up control of phytoplankton, which means that we did not consider the influence of zooplankton and higher trophic levels.
Let $P_i(z, t)$ denote the density of species $i$ at time $t$ and depth $z$. Assume that there are two limiting factors: the concentration of a nutrient, $N(z, t)$, and the light intensity $I(z, t)$, both of which are a function of the vertical position $z$.
Coupling of the nutrient and the phytoplankton dynamics leads to the following system of reaction-advection-diffusion equations
(Radach and Maier-Reimer 1975; Jamart et al. 1977; Klausmeier and Litchman 2001; Huisman et al. 2006)
%;
\begin{eqnarray}
\frac{\partial P_i}{\partial t} &=& \mbox{growth} - \mbox{loss} - \mbox{sinking} + \mbox{mixing}
\label{eq:1sppl}\\
  &=& \mu_i P_i - m P_i - v \frac{\partial P_i}{\partial z} +
\frac{\partial }{\partial z} \left[ D(z) \frac{\partial P_i}{\partial z}\right]
 \nonumber \\
 \nonumber \\
\frac{\partial N}{\partial t} &=&
 -\mbox{uptake}  + \mbox{recycling} + \mbox{mixing}  \label{eq:1spntr} \\
&=& -\alpha \sum_{i=1}^n \mu_i P_i + \varepsilon \alpha m \sum_{i=1}^n P_i  +
\frac{\partial }{\partial z} \left[ D(z) \frac{\partial N}{\partial z}\right]   \nonumber
\end{eqnarray}
where $\mu_i(N, I)$ describes the local growth rate of species $i$, $m$ is the mortality, $v$ is the phytoplankton sinking velocity, $D(z)$ is the depth dependent turbulent diffusivity, $\alpha$ is the nutrient content of a phytoplankton cell and $\varepsilon$ is the phytoplankton recycling coefficient.

Equations \eqref{eq:1sppl} and \eqref{eq:1spntr} are coupled by means of the growth rate $\mu_i(N, I)$ which depends on the local resource availability and also controls the nutrient uptake. Assuming that the limitation of growth follows the Monod kinetics (e.g. Turpin 1988) and both resources are essential (von Liebig's law of minimum), we obtain
\begin{equation}
\mu_i(N, I) = \mu_{\rm max}  \min \left( \mfrac{N}{H_N^{(i)} + N} , \mfrac{I}{H_I^{(i)} + I}
\right) \ ,
\label{eq:grrate}
\end{equation}
where $\mu_{\rm max}$ is the maximum growth rate, and $H_N^{(i)}$ and $H_I^{(i)}$ are the corresponding half saturation constants. Varying $H_N$ and $H_I$ allows to model, for instance, a species which is better adapted for light (a smaller $H_I$) or for the nutrient (a smaller $H_N$).

Light dissipates with depth as it is absorbed by the biomass, water, clay particles and many other absorbing substances. Assume that the light intensity decreases exponentially according to Lamber-Beer's law (see e.g. Shigesada and Okubo 1981, Kirk 1994)
\begin{equation}
I(z) = I_{in} \exp \left[ -K_{bg} z - k \int^z_0 \sum_{i=1}^n P_i(\xi, t) d \xi \right] \ ,
\label{eq:1splight}
\end{equation}
where $I_{in}$ is the incident light intensity, $K_{bg}$ is the turbidity of the water without biomass, and $k$ is the phytoplankton light-absorption coefficient.

To describe the water column stratification, we assume that the diffusivity $D(z)$ takes a large value $D_U$ in the upper layer and a much smaller value $D_D$ in the deep layers. A gradual transition from one area to another can be written in terms of a generalized Fermi function 
\begin{equation}
D(z) = D_D + \mfrac{D_U  - D_D}{1 + e^{(z-Z_{mix})/w}} \ ,
\label{eq:Dz}
\end{equation}
where $Z_{mix}$ describes the depth of the UML and the parameter $w$ characterizes the width of the transient layer.
In all numerical experiments 
%(with the exception of Fig.~\ref{fig:diff_inf}) 
we chose $D_U = \rm{50\ cm^2/s}$,
modeling well-mixed waters in the UML, and  $D_D = \rm{0.1...1.0 \ cm^2/s}$
for the lower layers (Lewis et al. 1986; Saggio and Imberger 2001; Smyth et al. 2001; Finnigan et al. 2002). 
The model parameters were chosen to describe clear ocean water (Huisman et al. 2006) and can be found in Table~\ref{tab:1}.

We simulate the model up to a depth of $z=Z_B$ and
assume impenetrable boundaries at the surface and at the bottom for the phytoplankton 
\begin{equation}
\left. \left( v_i P_i - D_z \mfrac{\partial P_i}{\partial z}\right) \right|_{z=0, Z_B} = 0 \ .
\label{eq:bound_pl}
\end{equation}
Since at the bottom of a sufficiently deep water column (we assume $Z_B = 300 {\rm \ m}$) the phytoplankton biomass vanishes together with flux, one can also use the Dirichlet boundary condition at the bottom $P_i(Z_B)=0$ and this will not influence our results. For the nutrient distribution we assume an impenetrable surface and a constant concentration at the bottom 
%and an impenetrable surface and a constant nutrient concentration at the bottom for the nutrient
\begin{equation}
\left. \mfrac{\partial N}{\partial z} \right|_{z=0} = 0, \qquad N(Z_B) = N_B \ .
\label{eq:bound_ntr}
\end{equation}

As initial condition we assumed a small uniformly distributed
concentration of phytoplankton ($ P(z, 0) < 1 {\rm \ cell \ m^{-3}}$), whereas for the nutrient we used two different initial profiles, describing a nutrient saturated water column ($N(z, 0) = N_B$) and a nutrient depleted upper layer ($N(z, 0) = 0 $ if $z \le Z_{mix}$). We have also explored the influence of other initial conditions, however we did not find any new solutions beside the ones described.  
Checking the stability of solutions, in each case we simulated dynamics for 50,000 system days (approximately 130 years).

The model was integrated using a backward difference method, based on the finite volume scheme (Pham Thi et al. 2005). For the numerical solution we have discretized all variables on a grid which consisted of 600 points.
Diffusion terms were approximated by a second order central discretization
scheme, the advection term was represented by a third-order upwind biased
formula, integration was made via the trapezoidal rule. The resulting system of
ordinary differential equations was solved by the CVODE package
(http://www.netlib.org/ode).
For model validation we have compared our simulation results with already published results (Huisman et al. 2006) and further verified that the results remain unchanged if we double the number of points in the grid. Furthermore, in some limiting cases it was possible  to compare our simulation results with analytic solutions.

\section{The single species model \label{sec:onesp}}

\begin{figure*}[tb]
\begin{center}
\centerline{\includegraphics[width=\textwidth]{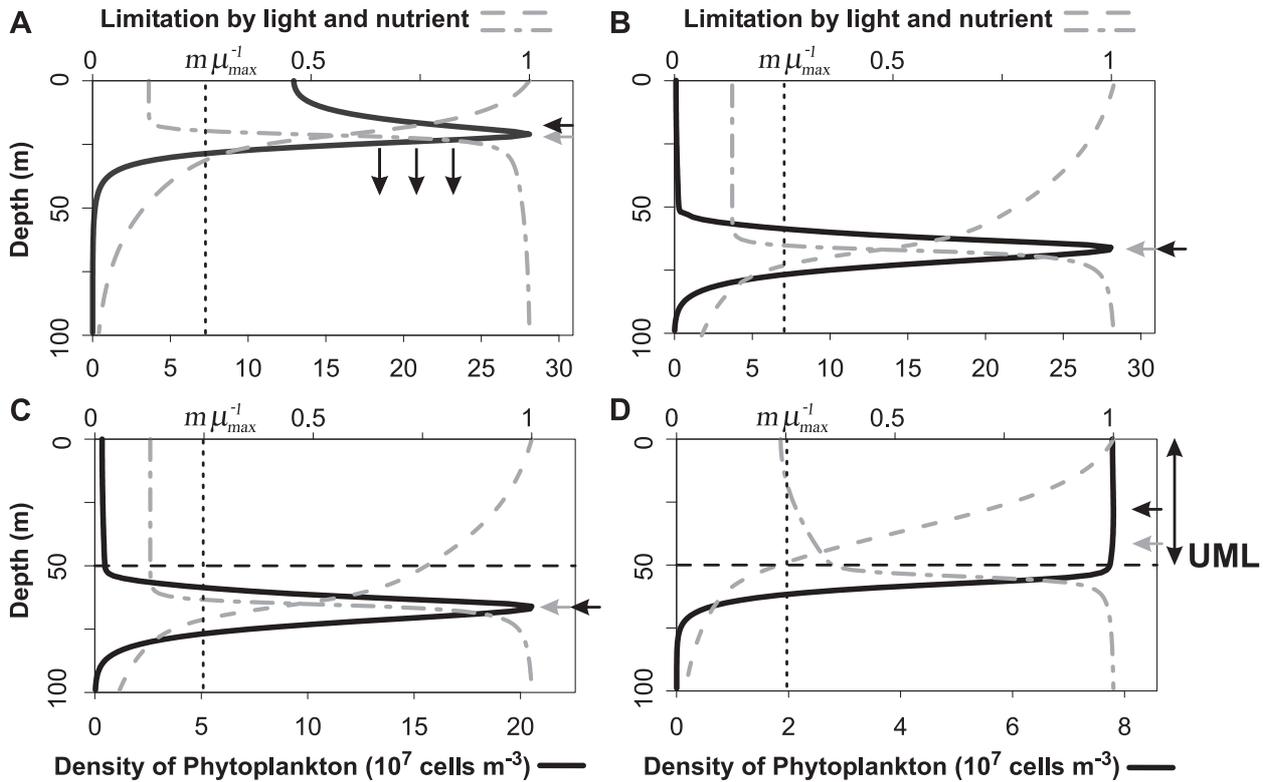}}
\caption{ Typical vertical phytoplankton profiles  in the single species model
for a system without a UML (top) and with a UML (bottom, depth $Z_{mix}$ indicated by the horizontal dashed line).  Plotted are the density of phytoplankton as function of depth, $P(z)$ (black solid line), and the growth limiting terms with respect to light, $I/(I+H_I)$ (gray dashed line), and to the nutrient, $N/(N+H_N)$ (gray dot-dashed line);
positive growth is possible where both regulating terms are larger than $m/\mu_{\mathrm{max}}$
(level of zero net growth indicated by the vertical dotted line).
Black and gray arrows show the centers of biomass $Z_m$ and net production $Z_g$, respectively.
({\bf A}) without a UML, a non-stable phytoplankton maximum close to the surface  is driven downwards (indicated by the arrows) and evolves to the stable state ({\bf B}).
Under the same conditions in a system with a UML, we observe two alternative stable configurations:
({\bf C}) a phytoplankton profile with a maximum in the deep layers (DCM),
or ({\bf D}) a profile with a maximum in the upper layer (UCM).
%, where $Z_m$ and $Z_g$ are decoupled.
}
\label{fig:centers}
\end{center}
\end{figure*}

%Typical vertical phytoplankton profiles  in the single species model for a system without a UML (top) and with a UML (bottom).  ({\bf A}) without a UML, a non-stable phytoplankton maximum close to the surface  is driven downwards (indicated by the arrows) and evolves to the stable state ({\bf B}).  Under the same conditions in a system with a UML, we observe two alternative stable configurations: ({\bf C}) a phytoplankton profile with a maximum in the deep layers (DCM), or ({\bf D}) a profile with a maximum in the upper layer (UCM), where $Z_m$ and $Z_g$ are decoupled.  Plotted are the density of phytoplankton as function of depth, $P(z)$ (black solid line), and the growth limiting terms with respect to light, $I/(I+H_I)$ (gray dashed line), and to the nutrient, $N/(N+H_N)$ (gray dot-dashed line), depth $Z_{mix}$ indicated by the horizontal dashed line; positive growth is possible where both regulating terms are larger than $m/\mu_{\mathrm{max}}$ (level of zero net growth indicated by the vertical dotted line).  Black and gray arrows show the centers of biomass $Z_m$ and net production $Z_g$, respectively.

\begin{figure*}[!hp]
\begin{center}
\linespread{1}
{\includegraphics[width=0.8\textwidth]{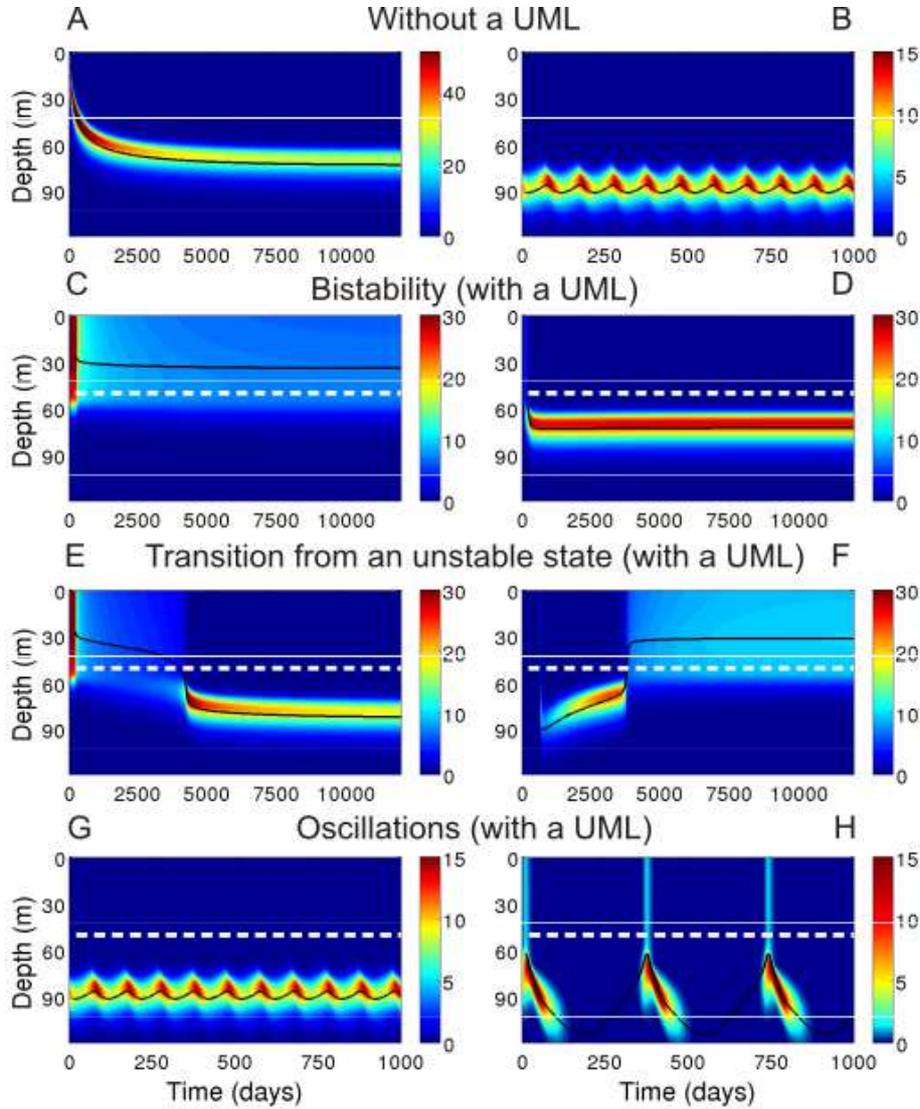} }
\caption{
Spatio-temporal evolution of typical phytoplankton profiles. Plotted is the
phytoplankton density $P(z,t)$ in color-coding ($\times 10^7$ cells $m^{-3}$) for different values of $D_D$ and initial conditions, in a system without a UML ({\bf A}-{\bf B}) and in the presence of a UML ({\bf C}-{\bf H}). Black lines track the evolution of the center of biomass, $Z_m$, white dashed lines indicate the depth of the UML.
Dynamics without a UML: ({\bf A}) gradual evolution of a DCM, $D_D = 0.3 \ \mathrm{cm^2/s}$, ({\bf B}) oscillations of the biomass for small mixing, $D_D = 0.12 \ \rm{cm^2/s}$.
Bistability (with a UML): depending on the initial conditions either ({\bf C}) a stable UCM or ({\bf D}) a stable DCM is formed (value of $D_D$ as in {\bf A}).
Transition from an unstable state: for any initial condition ({\bf E}) only a DCM is stable,  $D_D = 0.2 \ \mathrm{cm^2/s}$, or ({\bf F}) only a UCM is stable, $D_D = 0.4 \ \mathrm{cm^2/s}$, however the transient process may last a long time.
Oscillations of the biomass: ({\bf G}) oscillations are not affected by a UML, $D_D = 0.12 \ \mathrm{cm^2/s}$ (value of $D_D$ as in {\bf B}), or ({\bf H}) oscillations are induced by a UML, $D_D = 0.04 \ \mathrm{cm^2/s}$
(value of $D_D< D_{min} =  0.0408 \, \mathrm{cm}^2/s$).
}\label{fig:tdep}
\end{center}
\end{figure*}

We first concentrate on the dynamics of a single species population  and describe the formation of
a DCM in a water column without a UML.
Suppose that we start with an initially nutrient rich system ($N(z, t) = N_B$).
Thereby the nutrient limitation is negligible and  we observe a rapid formation of a transient phytoplankton maximum close to the surface (Fig.~\ref{fig:centers}A). This phytoplankton profile $P(z,t)$ is, however, not stable.
With the depletion of the nutrient in the surface layer the production layer, i.e. the layer where $\mu(N, I) \ge m$,
shifts downwards (see also Fig.~\ref{fig:tdep}A), until the system reaches a stable DCM configuration (Fig.~\ref{fig:centers}B). In this equilibrium the upward flux of nutrient compensates the nutrient consumption and any further sinking of the production layer is balanced by light limitation (Klausmeier and Litchman 2001).

Note that the spatio-temporal evolution of the concentration profile $P(z,t)$ depends on the growth conditions at all vertical positions. Due to the water turbidity and phytoplankton self-shading the light intensity $I(z,t)$ is reduced in deeper layers, thereby
increasing the limitation by light (dashed lines in Fig.~\ref{fig:centers}).
In contrast, the nutrient concentration $N(z,t)$ is close to zero within the bulk of phytoplankton biomass and above it, and increases almost linearly with depth below the phytoplankton peak
(see dot-dashed lines in Fig.~\ref{fig:centers} for the nutrient limitation of growth).
%and also Fig.~\ref{fig:steady_pl} in Appendix~\ref{sec:basicconf} for the resource distributions).
This shaping of the spatial dependence of the growth limiting factors self-consistently depends on the full phytoplankton density profile $P(z,t)$, a fact which makes the problem very hard to understand without mathematical simulation.

A rough insight into the time evolution of $P(z, t)$ can be gained by considering the centers of biomass $Z_m = \int z P dz / \int  P dz $ and of phytoplankton net production $Z_g = \int z g P dz / \int  g P dz $ (black and gray arrows in Fig.~\ref{fig:centers}). Here $g(z)$ describes the net phytoplankton production which includes growth, loss, sinking and mixing, i.e., the product $g P$ equals the right-hand-side of Eq. \eqref{eq:1sppl}.
In an incompletely mixed water column without a UML the position of
the center of mass, $Z_m$, follows that of $Z_g$. Thereby the phytoplankton production
around $Z_g$ shifts the mass center $Z_m$, changing the local nutrient
consumption and the light absorption, which in turn has an influence again on the position of the growth center $Z_g$. In this feedback-loop, the system eventually reaches a stable equilibrium configuration where both centers coincide, $Z_g = Z_m$, giving rise to a DCM (Fig.~\ref{fig:centers}B).

Now suppose that there is strong mixing in the upper layer.
If the bulk biomass is located sufficiently deep, then the mixing in the upper layer
has practically no effect and an identical DCM can persist (Fig.~\ref{fig:centers}C)
independent of whether or not a UML is present.
Note that we always assume that the depth of the  UML is smaller than the compensation depth
(i.e., a depth at which $\mu(I) =m$ in the absence of biomass,
see e.g. Sverdrup 1953, Huisman et al. 1999b).

By contrast, if the phytoplankton biomass is initially located close to the surface, it will be almost uniformly distributed within the UML. The position of $Z_m$ is then fixed approximately in the middle of the UML and is almost uncoupled from $Z_g$
(Fig.~\ref{fig:centers}D). Therefore a gradual shift of the center of mass into the deep layers is
no longer possible and the transition to a DCM can only take place if the light intensity below the
UML is sufficiently large to provide positive net growth in the deep layers -- otherwise the
phytoplankton remains trapped in the UML. We denote this stable configuration of a nearly uniform
phytoplankton profile in the UML as an upper chlorophyll maximum (UCM, note that all acronyms are listed in Table~\ref{tab:2}). Figs.~\ref{fig:centers}C and \ref{fig:centers}D show that in the presence of a UML, depending on the initial conditions, the system can undergo two very different spatial configurations of either a deep or an upper phytoplankton maximum (Yoshiyama and Nakajima 2002).

Fig.~\ref{fig:tdep} depicts the typical spatio-temporal evolution of the phytoplankton density.
Without a UML (Fig.~\ref{fig:tdep}A), an initial phytoplankton maximum at the surface slowly moves downward until the distribution converges to a stable DCM equilibrium.
However, as is shown in Fig.~\ref{fig:tdep}B, DCM's are not necessarily steady states. For small values of diffusivity $D_D$ the sinking of biomass may destabilize the density profile, yielding sustained regular or chaotic oscillations of biomass in the deep layer (Huisman et al. 2006).
Moreover, if the diffusivity is lower than the minimal persistence threshold
\begin{equation}
D_{min} = \frac{v^2}{4 (\mu(N_B, I_{in}) - m)}
\label{eq:Diff_min}
\end{equation}
the sinking phytoplankton population cannot survive and goes extinct (Riley et al. 1949; Shigesada and Okubo 1981).
Our numerical simulations show that in a system without a UML these three model outcomes (i.e., stationary DCM, oscillating DCM or extinction) are the only possible long-term solutions. Further, these solutions are globally attracting, which means that they are reached from any initial condition.

In the presence of a UML these solutions can be found as well, however the dynamics may be more complicated. Most notably, as already mentioned, the system may be bistable: under the same parameters as in Fig.~\ref{fig:tdep}A, an initially nutrient saturated water column gives rise to a UCM (Fig.~\ref{fig:tdep}C), whereas an initially nutrient depleted water column leads to a DCM (Fig.~\ref{fig:tdep}D). Note that we observe bistability only in a certain parameter region in which a DCM is not affected by the upper layer and a UCM contains enough biomass to limit growth in the deep layers. The diffusivity in the deep layers, $D_D$, is a suitable bifurcation parameter as it controls the nutrient flux from the bottom and an increase of $D_D$ rises the level of the DCM (Klausmeier and Litchman 2001).
Thus decreasing $D_D$ we obtain steady (Figs.~\ref{fig:tdep}E) or oscillatory DCM's (\ref{fig:tdep}G and \ref{fig:tdep}H), whereas larger values of $D_D$ yield UCM solutions (Fig.~\ref{fig:tdep}F).

Close to the bistability range we observe dynamics at very different time scales. Consider Fig.~\ref{fig:tdep}E which shows the transition from an unstable upper maximum to a stable DCM. Here, the simulation was started with a nutrient saturated water column which initially gives rise to a uniform phytoplankton profile in the UML.
Even though for the given parameter range the DCM is globally attracting, the center of biomass $Z_m$ at first is trapped inside the UML as described above.
This  configuration can be sustained for a rather long duration. However, with ongoing nutrient depletion the phytoplankton density is slowly declining and as soon as the biomass in the UML is not sufficient to shade light below it, the system undergoes a rapid transition to a DCM.
Similar dynamics at two very different time scales are observed in Fig.~\ref{fig:tdep}F, where a DCM gradually moves upward until it reaches the bottom of the UML and then the biomass rapidly shifts into the upper layer.
This transition from one solution type to the other can be very fast. It occurs on the biological time scale $\tau_B = \mu_0^{-1}$  and takes approximately 10-50 days for the model in our parameter range.
By contrast, the slow unstable dynamics before the transition is determined
by the diffusive time scale $\tau_B = (Z_B-Z_{mix})^2/(2D_D)$, which theoretically can last several years in deep waters. Obviously, in systems with essential seasonal variability
this transition will never be reached and the observed distribution will depend on the initial distribution of resources.

The bottom panel in Fig.~\ref{fig:tdep} shows oscillatory phytoplankton profiles in a system with UML. If the oscillatory state is deep enough then they are not affected by the presence of a UML (compare Figs.~\ref{fig:tdep}B and \ref{fig:tdep}F).
Interestingly however, the UML can support these oscillations even if $D_D < D_{min}$, i.e.
in a system where the population would die out without a UML (Fig.~\ref{fig:tdep}H).
The reason can be found in the fact that locally, within the well-mixed UML, diffusivity is much larger than the persistence threshold, $D_U \gg D_{min}$.
Now at the begin of the cycle, with practically no biomass the nutrient can freely diffuse towards the UML, where the population can outgrow sinking as soon as the nutrient concentration has reached a critical level. Then, if the biomass in the UML is not limited by light, the depletion of the nutrient shifts the production layer downwards into the weakly mixed water, where diffusivity is below the persistence threshold. There the population sinks further and
finally declines because of light limitation, so that a new portion of the nutrient can reach the UML and the cycle starts anew.
In this way the presence of a localized region with strong mixing at the top of the water column triggers an ``oscillatory pump'' for the biomass, with the effect that the persistence threshold of the sinking population disappears.
This is remarkable because most of the time in the cycle the bulk of the biomass remains located in the weakly mixed lower layers.

\begin{figure*}[!htb]
\begin{center}
\includegraphics[width=\textwidth]{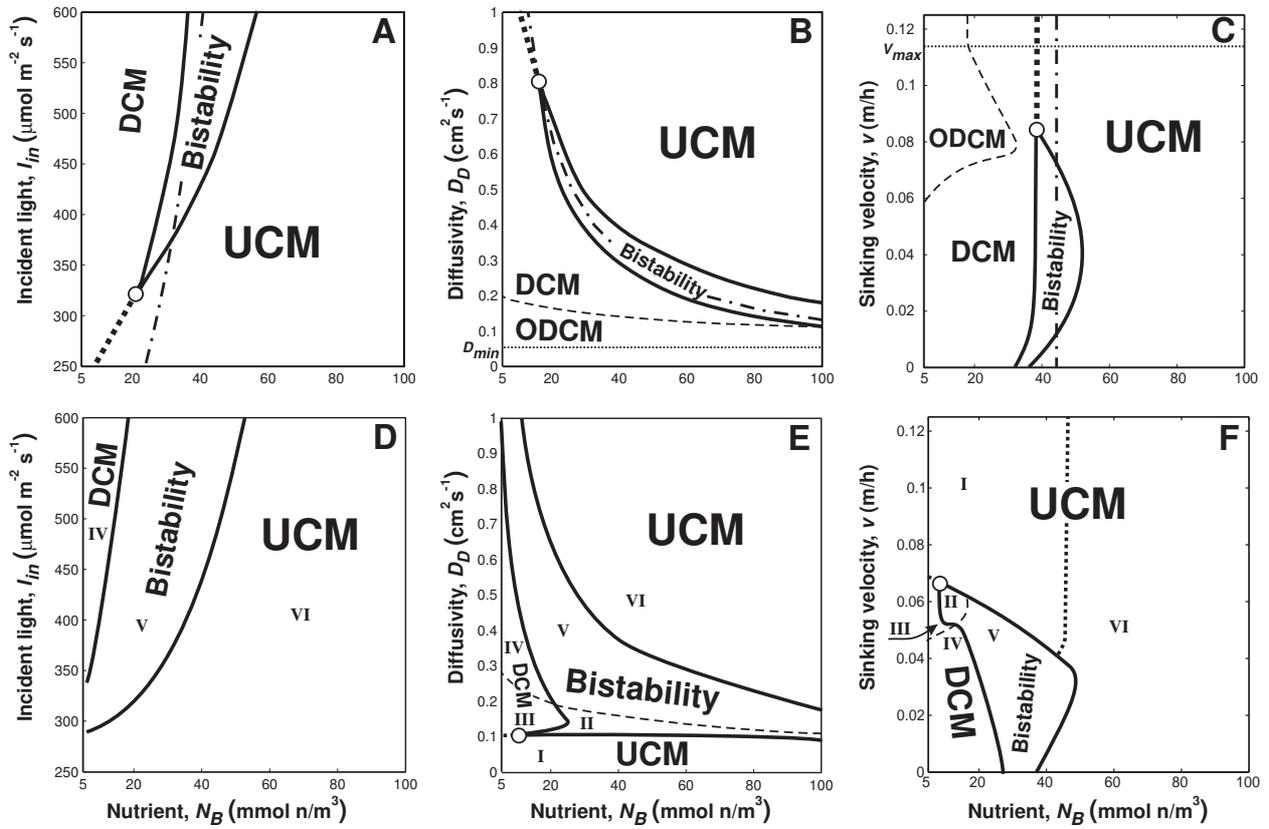}
\caption{Stability diagrams in the presence of a UML.
Shown are the regions of different dynamic outcome for a model with the  $N$-species solely (top)
and for a competition between $N$- and $I$-species (bottom) in the $(N_B, I_{in})$, $(N_B, D_D)$ and $(N_B, v)$ parameter planes. Solid lines separate regions with UCM, DCM or bistable dynamics.
Beyond the critical point (small circle) there is no clear separation between upper and deep biomass maxima.
Thick dotted lines in this regime indicate parameter values for which half of the biomass is distributed within and half below the UML.
Dashed lines separate regions of stationary and oscillatory solutions.
Dash-dotted lines in (\textbf{A})-(\textbf{C}) show the analytic border of stability of a UCM, see Eq.~\eqref{eq:LimLight}.
Thin horizontal dotted lines in (\textbf{B}) and (\textbf{C}) indicate the persistence threshold \eqref{eq:Diff_min}
in an equivalent water column without a UML.
Roman numerals enumerate different dynamic regimes in the two-species model: I and VI -- UCM, II -- bistability between ODCM and UCM, III -- ODCM, IV -- stationary DCM, V -- bistability between  stationary DCM and UCM.
The gray thick dotted line in (\textbf{F}) shows the border between regions I and VI.
}
\label{fig:2sp}
\end{center}
\end{figure*}

To investigate the system behavior in a large range of parameters we performed
simulations for 900 pairs of $(N_B, I_{in})$, $(N_B, D_D)$ and $(N_B, v)$.
The results are presented in the stability diagrams Figs.~\ref{fig:2sp}A - \ref{fig:2sp}C. As shown in
Fig.~\ref{fig:2sp}A, large values of $I_{in}$ in general lead to a DCM and large $N_B$ to a UCM, while for
intermediate resource levels we observe a region with bistable behavior. The bistability range is reduced for smaller values of $I_{in}$ or $N_B$ and disappears at a critical point ($I_{in} \approx
350 \  \rm {\mu mol \ photons \ m^{-2} \ s^{-1}}$ and $N_B \approx 25 \ \rm{mmol} /m^3 $ in Fig.~\ref{fig:2sp}A).
Note that the lower border of the bistability range is quite well described by the analytical criterion Eq.~\eqref{eq:LimLight},
which is derived from the condition that the phytoplankton net production rate below the mixed layer is not positive (dash-dotted lines in Fig.~\ref{fig:2sp}).
Interestingly, outside the region of bistability close to the critical point
smooth transitions from deep to surface maxima are possible. In this case intermediate density profiles with no clear separation between DCM and UCM appear and the phytoplankton biomass is located in both parts of the water column.
%(see Fig.~\ref{fig:steady_pl} in Appendix \ref{sec:basicconf}).
To visualize the distribution of biomass in this regime,
parameter combinations, for which the median of the biomass distribution is located at $Z_{mix}$ (i.e.,
half of the biomass is distributed within and half below the UML),
are indicated as thick solid lines in Fig.~\ref{fig:2sp}.

Fig.~\ref{fig:2sp}B demonstrates the bistability range in the $(N_B, D_D)$ parameter plane, where the bifurcation lines have an almost hyperbolic form. A UCM appears for large values of $D_D$ and $N_B$, whereas small values favor a DCM. Very small values of $D_D$ result in oscillatory DCM (ODCM) solutions, in accord to the results of Huisman et al.~(2006), and due to the presence of a UML this behavior can still be observed, and the population does not go extinct, for $D_D< D_{min}$.

In the $(N_B, v)$ parameter plane the bifurcation lines have almost vertical structure (Fig.~\ref{fig:2sp}C). Most notably, the minimal value of $N_B$ which can lead to upper biomass maxima practically is independent from $v$.
Thus sinking has no strong influence on the stationary solutions in a well mixed layer.
By contrast, oscillatory solutions arise only for a large enough sinking velocity.
Again, the population can persist beyond the persistence threshold $v>v_{max}$, where $v_{max}$ is the maximal sinking velocity which allows the survival of a population in a system without a UML. $v_{max}$ can be obtained from Eq.~\eqref{eq:Diff_min} for a given diffusivity.

\section{The two species model \label{sec:2sp}}

We now turn to the influence of a UML on
the competition of two species which are differently adapted to the conditions at the surface or in the deep water. For this, we extend the model to contain two phytoplankton species which differ in their respective half saturation constants $H_I$ and $H_N$ in Eq.~\eqref{eq:grrate}. More specifically, we consider the competition between a most light-limited species ($I$-species) characterized by a low $H_N$ and a large $H_I$ value and so is forced to live close to the light source, and a most nutrient-limited species ($N$-species) with low $H_I$ and large $H_N$ that will usually do better in deep water.
Note that the results of the previous section  hold for the $N$-species.

To avoid the influence of different growth, mortality and consumption rates, we study the simplest possible case and keep the other species' parameters identical in both species (see Table~\ref{tab:1}). Thereby, in a well mixed uniform environment (e.g., a chemostat) these species would have parallel consumption vectors, so that the success of one species over the other depends only on the resource concentrations in the absence of biomass; with the consequence that in a chemostat such species cannot coexist and the outcome of their competition cannot be bistable (Tilman 1980, 1982).
In a well mixed water column, where the light intensity reduces with depth, the competition of these species is more complicated, however the same results still hold (Huisman and Weissing 1995); whereas
in an incompletely mixed water column, assuming only the light gradient, both species can coexist
in a narrow parameter region (Huisman et al. 1999).
In the following we show that the outcome of competition is drastically altered in the presence of two opposing resource gradients combined with a space dependent diffusivity, resulting in new regions of competitive exclusion, bistability and coexistence.

\begin{figure*}[!hp]
\begin{center}
{\includegraphics[width=\textwidth]{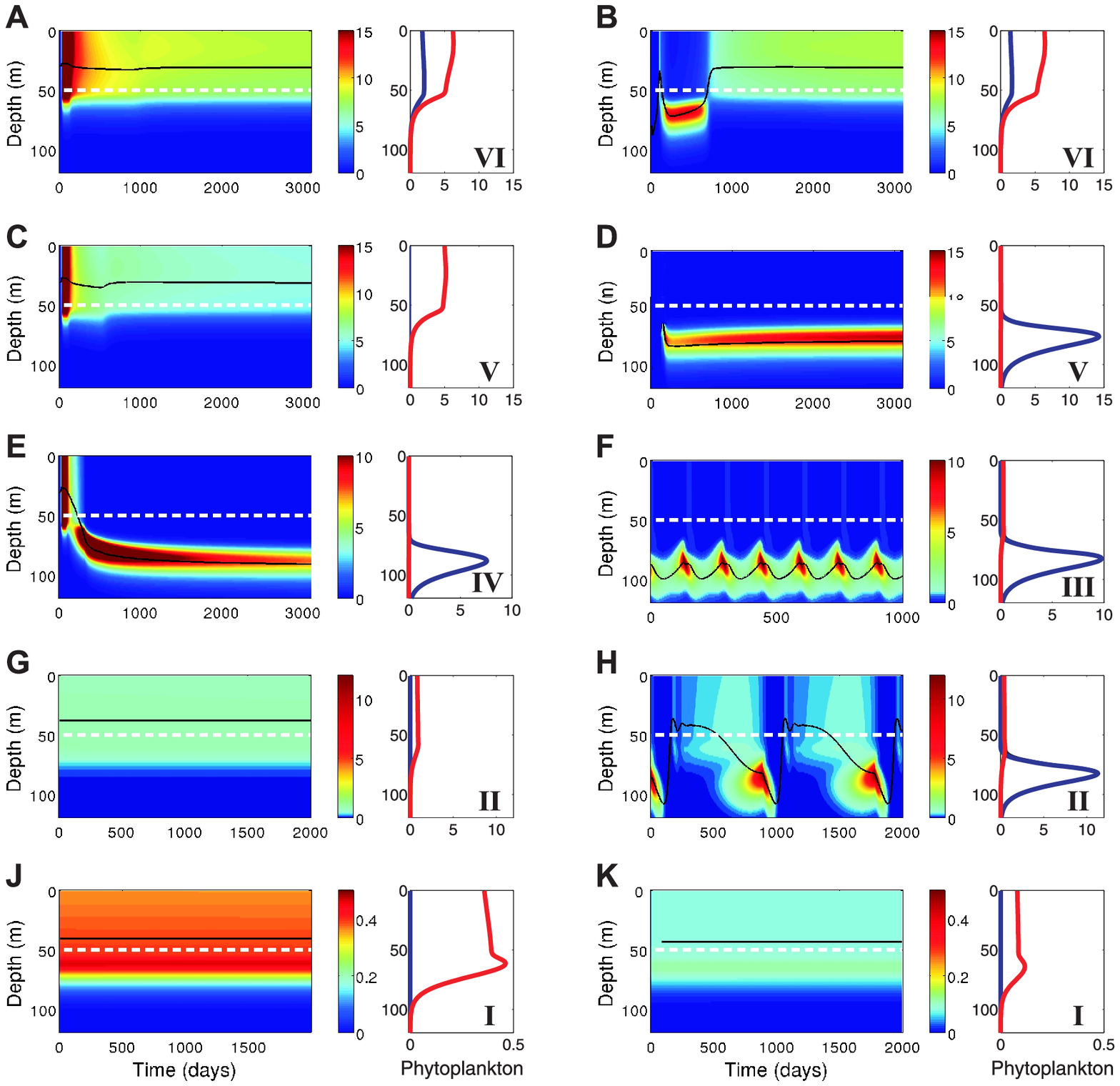} }
\caption{Typical phytoplankton profiles in the two species model in the presence of a UML.
Shown are the time evolution of the total phytoplankton density in color coding and the time averaged density profiles
of the I-species (red) and the N-species (blue) for different values of $D_D$ and initial conditions (roman numerals indicate the corresponding dynamic regime).
(\textbf{A}) and (\textbf{B}) different initial conditions lead to the same outcome, coexistence of both species in a UCM, $D_D = 0.8 \  \mathrm{cm^2/s}$. (\textbf{C}) and (\textbf{D}) bistability between a UCM of the I-species and a DCM of the N-species, $D_D = 0.5 \  \mathrm{cm^2/s}$.
(\textbf{E}) and (\textbf{F}) $N$-species always wins and forms a stable or an oscillatory deep maximum, $D_D = 0.2 \ \mathrm{cm^2/s}$ and $0.15 \  \mathrm{cm^2/s}$, respectively.
(\textbf{G}) and (\textbf{H}) bistability between a stable UCM and an oscillatory DCM, $D_D = 0.10 \  \mathrm{cm^2/s}$. (\textbf{J}) and (\textbf{K}) stable UCMs, $I$-species always excludes $N$-species, $D_D = 0.05 \  \mathrm{cm^2/s}$ and $0.01 \ \mathrm{cm^2/s}$, respectively.
Black lines track the evolution of the center of biomass, $Z_m$, white dashed lines indicate the depth of the UML.
}\label{fig:tdep_twosp}
\end{center}
\end{figure*}

Figs.~\ref{fig:2sp}D - \ref{fig:2sp}F present the two-species stability diagrams in the different
parameter planes, listing all possible model outcomes.
Basically, the overall shape of the transition lines between bistable and UCM/DCM states remain identical to those of the one-species model.
The most notable difference is that the bistability range (regime V) has become much wider and is extended toward smaller values of $N_B$ (compare top and bottom panels in Fig.~\ref{fig:2sp}).
By contrast, the transition line
to the UCM remains largely unchanged, since it is mainly determined by the minimal depth of the deep maximum at which it is not affected by the UML. For our set of parameters, the deep maximum is always formed by the $N$-species (whereas the upper maximum can be formed by both species) and so this boundary does not change with addition of the $I$-species.
Only in Fig.~\ref{fig:2sp}F this boundary is essentially altered, reflecting the fact that sinking influences much stronger the species which occupies the deep weakly mixed layers.

More interestingly, the two species system exhibits a variety of patterns and new dynamical regimes (enumerated by roman numerals I - VI in Figs.~\ref{fig:2sp}E and \ref{fig:2sp}F).
A detailed representation of the biomass dynamics and phytoplankton profiles in each dynamic regime can be found in Fig.~\ref{fig:tdep_twosp}.

\begin{figure*}[!htb]
\begin{center}
\includegraphics[width=\textwidth]{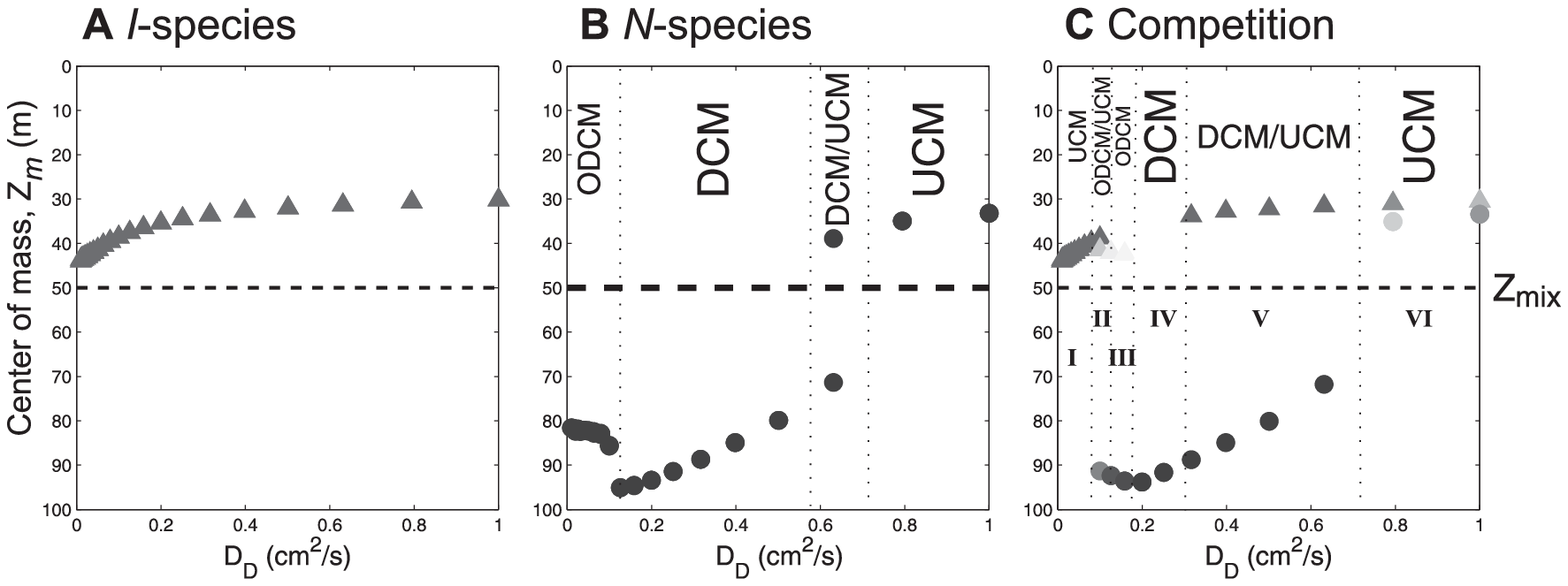}
\caption{Vertical cross section for $\rm{N_B = 20\  mmol/m^3}$
through the $(N_B, D_D)$ parameter plane of Figs.~\ref{fig:2sp}B and E, showing the center of mass, $Z_m$, of the $I$-species (triangles) and the $N$-species (circles) as a function of $D_D$.
(\textbf{A}) Monoculture of the $I$-species,
(\textbf{B}) monoculture of the $N$-species,
(\textbf{C}) two-species system.
In (\textbf{B}) and (\textbf{C}) several dynamic regimes (separated by vertical dotted lines) can be distinguished (see text). The competition outcome in (\textbf{C}) : I -- $I$-species wins, II -- bistability:
either $I$-species wins or the species coexist due to oscillations, III -- coexistence due to oscillations, IV -- $N$-species wins, V -- bistability, VI -- coexistence. In the regimes of coexistence, the fraction of biomass is indicated by the intensity of gray color.
}
\label{fig:pos_pl}
\end{center}
\end{figure*}

These transitions and the species composition can be visualized in more detail by tracing the depth of the center of biomass as a function of $D_D$. This is shown in Fig.~\ref{fig:pos_pl} for the single species and the two species case. This figure adds a third dimension to the vertical cross section through Figs.~\ref{fig:2sp}B and \ref{fig:2sp}E at $\rm{N_B = 20\  mmol/m^3}$.
While the bulk of the biomass of the $I$-species monoculture is always located within the UML (Fig.~\ref{fig:pos_pl}A), a monoculture of the $N$-species for decreasing values of $D_D$ undergoes four dynamical regimes, as described in the previous section, namely:  UCM, UCM/DCM bistability, stationary DCM and ODCM (Fig.~\ref{fig:pos_pl}B).

By contrast, the competition of two species  leads to more intricate behavior, as is demonstrated in Fig.~\ref{fig:pos_pl}C.
At the high end of the $D_D$ range both species are located in the UML and can either coexist (region~VI, Figs.~\ref{fig:tdep_twosp}A and \ref{fig:tdep_twosp}B) or the $N$-species will competitively exclude the other species if light is the only limiting factor (not shown).
With reduction of $D_D$, the $N$-species can form a DCM in the lower layers, yielding
a large bistability range between a DCM of the $N$-species and a UCM of the $I$-species (region~V, Figs.~\ref{fig:tdep_twosp}C and \ref{fig:tdep_twosp}D). Thereby, the bistability of the phytoplankton profiles goes together with a  bistability in the competition outcome.
Following reduction of $D_D$ leads to competitive exclusion of the $I$-species and only stable DCMs of the $N$-species are found (region~IV, Fig.~\ref{fig:tdep_twosp}E).

For even smaller values of $D_D$ the stationary DCM loses its stability yielding an oscillatory DCM (region~III, Fig.~\ref{fig:tdep_twosp}F). Interestingly, in this regime, the $I$-species may obtain a time window during each cycle to establish a population in the UML (slightly visible as light blue stripes in Fig.~\ref{fig:tdep_twosp}F), but the next rapid outburst of the $N$-species will again lead to the dominance of the $N$-species.
With further reduction of $D_D$, the period of the DCM oscillations increases until finally the $N$-species cannot outcompete an established population of the $I$-species in the upper layer. However the oscillatory solution, once established, can persist for the same parameter values.
Thereby, we observe a second bistability regime, of either a stationary UCM of the $I$-species
or the coexistence of both species due to oscillations
(region~II, Figs.~\ref{fig:tdep_twosp}G and \ref{fig:tdep_twosp}H).

Finally reducing $D_D$ to very small values leads to a surprising result:
the  $I$-species always wins the competition and the steady UCM formed by the $I$-species
becomes the single possible attractor in the system, independent of the initial conditions (region~I, Fig.~\ref{fig:tdep_twosp}J).
Thus, a strongly light limited species is able to establish a steady population in the UML, outcompeting a less light limited species which, in the absence of the former, would form an oscillatory maximum in the deep weakly mixed layers.
%Note, that an increase of the sinking velocity leads to the same result (region I in \ref{fig:2sp}F).
Note that the $I$-species survives even  if $D_D < D_{min}$, provided that the nutrient concentration in the upper layer is sufficiently high (Fig.~\ref{fig:tdep_twosp}K).

%\section{Comparison of the competition outcomes in the system with and without a UML \label{sec:without}}
\begin{figure*}[!tb]
\begin{center}
{\includegraphics[width=\textwidth]{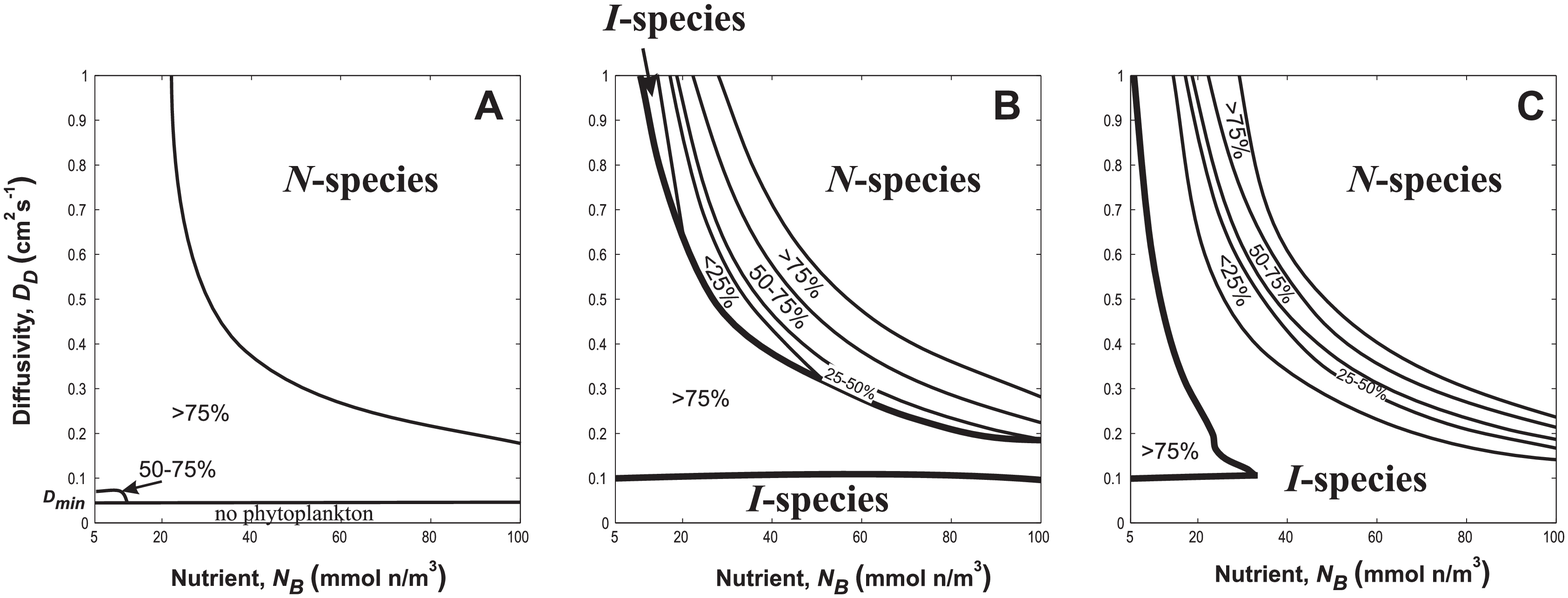}}
\caption{Comparison of the competition outcomes in a system without a UML (\textbf{A}) and with a UML (\textbf{B}) and (\textbf{C}). Figures (\textbf{B}) and (\textbf{C}) show the final species composition when the UML was initially nutrient depleted (favors to a DCM) or saturated (favors to a UCM), respectively. Shown are the ranges where either $I$-species or $N$-species wins. In the ranges where two species coexist the fraction of the $N$-species is presented. The solid lines are those transition lines which coincide with the bifurcation lines in Fig.~\ref{fig:2sp}E, i.e., when the change in the species composition is accompanied by a change in the vertical profile.}
\label{fig:comp}
\end{center}
\end{figure*}

Fig.~\ref{fig:comp} shows the outcome of competition between two species in the $(N_B, D_D)$ coordinate plane. In the system without a UML (Fig.~\ref{fig:comp}A) the $N$-species either wins or dominates (more than $75\%$ of the biomass). Furthermore, we observe no phytoplankton for diffusivity below $D_{min}$. Comparison of Fig.~\ref{fig:comp}A with Figs.~\ref{fig:comp}B and \ref{fig:comp}C show that a UML plays a crucial role improving the competitive abilities of the $I$-species, which occupies a shallower depth because of its higher light requirements. Shifts in the competition outcome can be caused by a change in the resource availability or by a bifurcation in the vertical biomass distribution. The thick lines in Fig.~\ref{fig:comp}B and \ref{fig:comp}C indicate those transitions in the species composition which are caused by a change of vertical biomass profiles. These lines coincide with the bifurcation lines in Fig.~\ref{fig:2sp}E. In the presence of a UML the $I$-species can win under the following conditions: (i) when the diffusivity in the lower layers is reduced ($D_D<0.1 \rm{cm^2/s}$) and the $N$-species alone would exhibit oscillating behavior; (ii) in the range of bistability, where either a UCM of $I$-species or a DCM with a large fraction of $N$-species appears; (iii) in the UML for sufficiently strong $D_D$ and low $N_B$ ($~15\  \rm{mmol \ n/m^3}$ in Fig.~\ref{fig:comp}B), when the $N$-species solely would still occupy the UML, but becomes a weaker competitor under stronger nutrient limitation. We also note that change of the vertical profiles causes non-monotonous shifts in the species composition.

\section[Graphical approach]{Graphical approach to analyze the competition outcome \label{sec:gr}}

\begin{figure*}[!htb]
\begin{center}
\includegraphics[width=\textwidth]{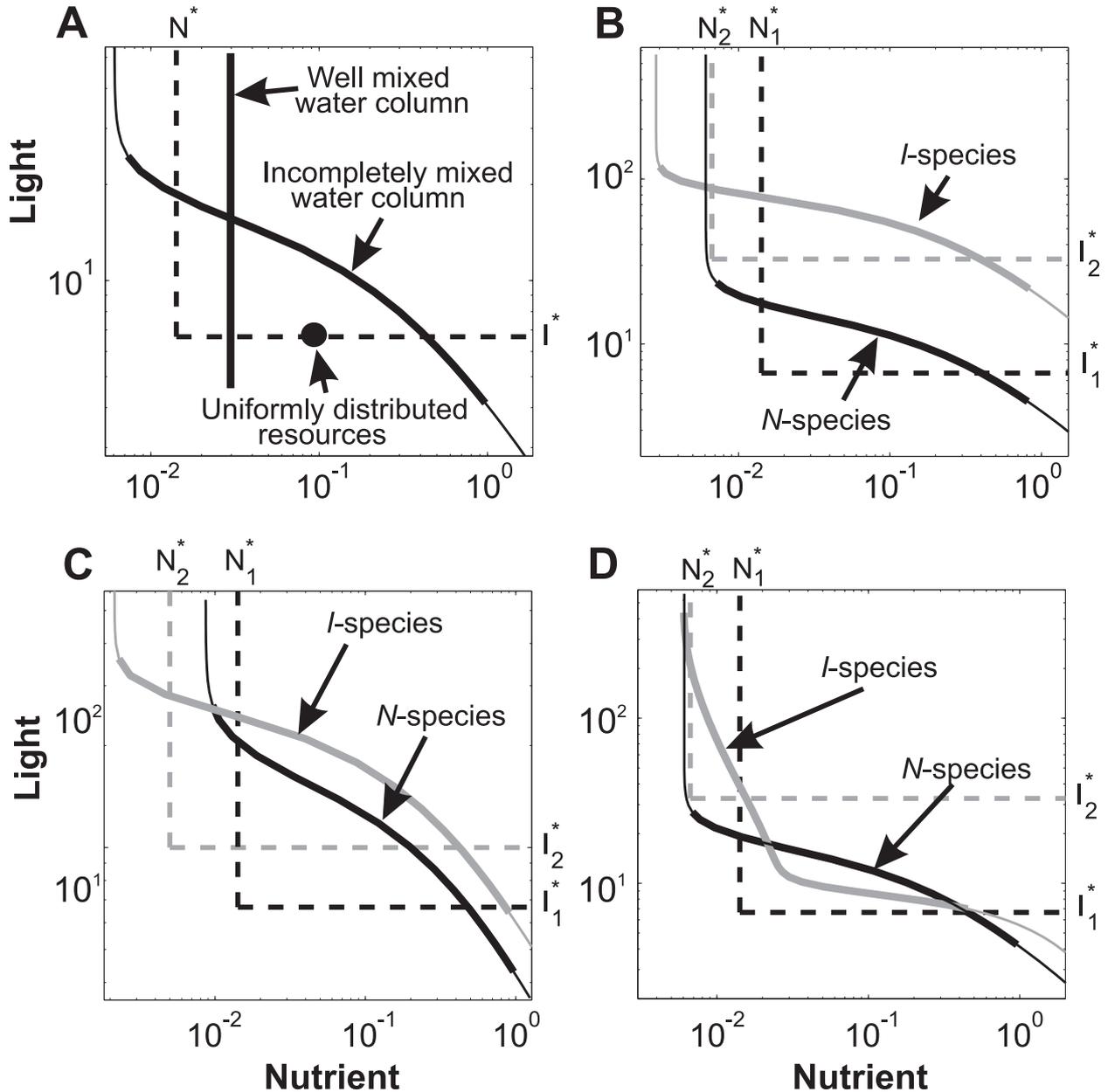}
\caption{Graphical approach for analyzing the competition outcome.
%in a spatially extended system.
Plotted are zero net growth isoclines (dashed lines) and system state curves in equilibrium (SSC, solid lines) for monocultures of the $N$- and the $I$-species, black and gray lines, respectively.  
(\textbf{A}) Comparison of differently mixed systems.
(\textbf{B}) Competitive exclusion ($N$-species wins) and (\textbf{C}) coexistence in a system without a UML.
(\textbf{D}) Bistability under the same parameters as (\textbf{B}) but in a system with a UML.
%Due to diffusion the phytoplankton density is larger then zero everywhere in the water column, 
For each SSC the area containing $90\%$ of the biomass is marked as a thick line. Note the logarithmic scale of the axes.
Parameter values: $N_\mathrm{B} = 30\  \mathrm{mmol}/\mathrm{m}^3$, $v=0$.
(\textbf{B, D}) $H_I=98\ \mu$mol photons m$^{-2}$  s$^{-1}$, $H_N=0.02$ mmol nutrient m$^{-3}$ for the $I$-species;
(\textbf{C}) $H_I=30\ \mu$mol photons m$^{-2}$  s$^{-1}$, $H_N=0.013$ mmol nutrient m$^{-3}$ for the $I$-species.
} \label{fig:isoclines}
\end{center}
\end{figure*}

%As we have shown in the previous section, 
In a spatial system the outcome of competition is hard to deduce from first principles. 
% This can be related to the fact that 
First, the production layers of differently adapted species typically will be located at different depths, which tends to reduce the strength of interspecific competition.
% but even more importantly, 
Second, the competition becomes indirect, 
% and so more difficult to understand, 
since the occurrence of phytoplankton at a certain depth can completely reshape the resource distributions in other parts of the water column.
To obtain an insight into the mechanisms of competition in such an environment, we suggest a graphical approach, which is based on a well known method developed by Tilman for mixed systems (1980, 1982).

If all resources and the biomass are uniformly distributed, the state of the system may be represented as a point in a multidimensional resource space (Fig.~\ref{fig:isoclines}A).
As the biomass grows and takes-up resources this system state point moves in the resource plane
(along the consumption vector) until it hits the zero net growth isocline, characterized by a balance between growth and loss processes.
From the condition $m=\mu_i(N,I)$ in equation~\eqref{eq:grrate} we can find such  values $N^\star$ and $I^\star$ so that the specific growth rate equals the mortality
\begin{equation}
  N^* =  \mfrac{m}{\mu_{max}-m}  H_N   \ , \quad I^* = \mfrac{m}{\mu_{max}-m} H_I 
\label{Eq:limN} \ .
\end{equation}
These values
%, which are just proportional to the species's half-saturation constants,
determine the location of the zero net growth isoclines (Fig.~\ref{fig:isoclines}).
The isoclines divide the resource plane into areas of positive and negative local population net growth and mark possible resource combinations in equilibrium. % so that the species with the least resource requirement survives.

In a spatially-extended system, where the resource values change with depth $z$,
%the competition outcome must be calculated for every vertical position. Consequently, 
the state of the system can be represented as a curve in the resource space, which is parametrically determined by the resource values $(N(z), I(z))$. In this sense, the notion of a system state point in a homogeneous system naturally extends to a system state curve (SSC) in a spatial explicit system.  A special case is given by a well mixed water column (Fig.~\ref{fig:isoclines}A): while the nutrients are uniformly distributed, the light intensity decreases exponentially with depth and the SSC reduces to a line segment $\{{N = \mbox{const},\ I_{out} \leq I \leq I_{in}}\}$, where $I_{out}$ is the light intensity at the bottom of the water column (Huisman  and Weissing 1995).

To simplify the discussion, in the following we always focus on the system state curve (SSC) in equilibrium.
Fig.~\ref{fig:isoclines}A shows a typical simulation outcome for the SSC of a single-species population in an incompletely mixed water column. Note that we use logarithmic scaling to magnify the location of the SSC close to the zero net growth isoclines.
%(see also Fig.~\ref{fig:SSC}). 
Due to the diffusive mixing the SSC does not settle at the zero net growth isoclines, as would be the case for a uniform distribution of the resources. Instead the SSC extends into the area of positive (``favorable range''), as well as into the area of negative population net growth (``unfavorable range''). To give a crude insight into the density variation along a SSC we indicate the central range of the curve which contains 90\% of the biomass as a thick line.% in Figs.~\ref{fig:isoclines}A-D.

For the analysis of a multi-species system, we find it convenient to calculate the SSC independently for each species in the absence of all other species. So each species attains its own SSC which would indicate its single-species resource configuration in equilibrium.
Fig.~\ref{fig:isoclines}B shows an apparent example, illustrating the case of competitive exclusion. In this example the SSC of the $N$-species lies below the null growth isocline of $I$-species.
Thus, the $N$-species reduces the resource levels at all vertical positions to values which are smaller than the resource requirements of the $I$-species and so do not permit positive net-growth of the invading $I$-species. 
%As a result the $I$-species cannot invade in the presence of the $N$-species.
By contrast, the $I$-species' SSC is located mostly above the null growth isocline of the $N$-species and so allows a positive net-growth of the $N$-species. As a consequence, in this example the $I$-species will always be excluded as it has larger resource requirements.

Fig.~\ref{fig:isoclines}C demonstrates an example of the coexistence of two species. In this figure both curves intersect in such a way, that the SSC of one species is below that of its competitor in an essential part of its favorable range, and vice versa. In these conditions both species can coexist because both are superior competitors at different depths. 
In general, however, if the SSC of one species lies above the zero null growth isocline of its competitor, there is no unambiguous answer to the question ``whether or not the latter species can invade?'' 
%On the one hand this species obtains a favorable patch, where $\mu>m$, and could invade in the limit of zero diffusivity (and zero sinking). But on the other hand, if $D>0$, it will invade only if this patch and the growth rate there are large enough to compensate for losses into unfavorable layers (see e.g. Ryabov and Blasius 2008). More precisely 
The possibility of invasion will depend on the principal eigenvalue of a reaction-diffusion equation, characterizing the growth of the biomass (Cantrell and Cosner 2001; Ryabov and Blasius 2008),  a problem which unfortunately can only be solved in some simple cases. %Therefore we find it more robust to analyze the intersection of the SSCs rather than their placement with respect to the isoclines.

%However, if the SSC of a species would lie above that of its competitor then the former would be excluded because it requires higher resources.

%Analysis of SSCs becomes a useful tool for understanding the role of a UML. 
%The intensive mixing in the upper layer changes the form of system state curves and they can intersect in more intricate ways. 
%Take for instance the situation shown in Fig.~\ref{fig:isoclines}D. 

As is shown in Fig.~\ref{fig:isoclines}D the analysis of SSCs can also help to gain 
%becomes a useful tool for 
insight into the role of a UML. Here we use the same set of parameters
as in Fig.~\ref{fig:isoclines}B, but in the presence of a UML. In this case the main part of the
$I$-species biomass is located close to the surface, while the $N$-species still forms a DCM. Hence the
presence of the UML leaves the SSC of the $N$-species unaffected while the SSC of the $I$-species
becomes steeper and is shifted in direction of smaller resource values.
%(see also Appendix~\ref{sec:slope}). 
This change in the form of the $I$-species' SSC results in new intersections of both system state curves with the result to improve the competing abilities of the $I$-species. 
While the presence of the $N$-species still completely prevents an invasion of the $I$-species, as in the case without a UML, now the SSC of the $I$-species lies below the main part of the $N$-species' SSC. 
%This means that the $I$-species reduces resource values to such values which hinder the growth of the $N$-species. 
In this example, the $N$-species can exist in the presence of the $I$-species, however its fraction is small so that its contribution to the total resource distribution is negligible. As a result, in the presence of a UML the outcome of competition is bistable and depends on initial conditions.

Our investigations show that the analysis of a species' SSC 
can further our understanding about resource competition 
%has a potential to provide an understanding of the dynamic behavior and competition 
in heterogeneous multi-species systems.  
%Especially in the important case, when the spatial size of the areas of positive net growth is equal for both species in monoculture, it is possible to precisely predict the outcome of competition. However we concede that, if the favorable patch sizes of the species are very different, it may be difficult or even impossible to analyze the competition relying solely upon the SSC.
What is more, one can obtain rigorous results for biologically similar species. A more detailed analysis will be presented in a forthcoming article (Ryabov and Blasius, in preparation).
% rigorous results can be obtained for biologically similar species (a more detailed analysis will be presented in Ryabov and Blasius, in preparation).

\section{Discussion}

In this article we investigated the influence of an upper mixed layer on the distribution and competition of phytoplankton species in a water column, in which inverse resource gradients (of light and a nutrient) can limit growth of the biomass. 
%In our model we study a bottom-up control of phytoplankton, which means that we did not consider the influence of zooplankton and higher trophic levels.
In this system the location of a production layer is not fixed, rather it depends on the initial and boundary conditions, on the species requirements, and on the stage of the process. 
Together with the presence of differently mixed areas (i.e., the UML and lower layers), this leads to a plethora of phenomena, including bistability of phytoplankton profiles, changes in the competition outcome, and new critical conditions for the survival of a phytoplankton population.
Thus our study not only proves the UML to be an important factor % that cannot be neglected in the description of a vertical water column, 
with the potential to shape the spatial distribution and species composition of phytoplankton, but also reveals insights of general ecological importance.

While previous theoretical investigations have usually focused on either a fully mixed or an incompletely mixed system, the presence of a UML requires a combination of these approaches.
Including both factors, by dividing the water column into two separate compartments, Yoshiyama and Nakajima (2002, 2006) showed the existence of bistability in the spatial distribution of a phytoplankton monoculture. Our work confirms and extends these findings on the following key points. First, assuming a gradual change of diffusivity with depth, our modeling approach integrates the whole water column into a single framework.
This allows us, for example, to investigate the influence of mixing in both the upper and the lower layers.
Secondly, we analyze the competition of species which are differently adapted to the availability of nutrients and light. As we show, this has drastic effects because the species composition strongly correlates with the spatial patterning.
And finally, our analysis includes the case of a stratified lower layer, when a stable DCM cannot persist and oscillatory or chaotic solutions appear, which again have a strong influence on the species competition in the system.

In comparison with lower layers three factors make a UML more favorable for phytoplankton species. Firstly, a UML has only one border with an unfavorable environment below the euphotic zone. In contrast, a deep production layer has two such boundaries and diffusion of cells upward and downward from it leads to additional losses due to either light or nutrient limitations. 
%This decreases the total loss rate in a UML in comparison with lower layer.
%Thus the loss rate from an UML into the hostile environment than that in the deep layers.
%Therefore the losses to unfavorablelayers are smaller than those in deep layers, where the unfavorable environment is located below and above the production layer. 
%Finally note that a UML has only one boundary with the unfavorable environment below the euphotic zone. In contrast a deep production layer has two such boundaries, as the diffusion upward and downward from it leads to additional losses due to either light or nutrient limitations. Thereby the loss rate from the UML into the hostile environment is roughly two times smaller than that in the deep layers. This effect also lowers the requirements for resources in the UML. 
Secondly, the strong mixing in the upper layer 
%always provides a sufficient propagation velocity, which 
allows for the survival of a sinking phytoplankton population. This additionally supports a population of phytoplankton even if the diffusivity in the lower layers becomes very small. Higher diffusivity also reduces the effective mortality rate of sinking phytoplankton species (see Appendix~\ref{sec:loss}).
% as a result, the critical resource values in a UML are also smaller than those in deeper layers. 
Thirdly, a UML promotes a nearly uniform distribution of nutrients, which makes the nutrient consumption
more efficient and gives an additional competitive advantage for a species inhabiting a UML.  
All these factors decrease the total loss rate
%, which can be interpreted as an effective mortality rate, as result 
and, hence, reduce the resource requirements.
However, we note that a deep and turbid UML can also play a negative role, as it can lead to the extinction of species due to light limitation (Huisman et al. 1999b).

%As one major outcome of our model, the location of the production layer depends on initial conditions and on the distribution of resources and phytoplankton. As a consequence of this mobility, the production layer can be steadily located either within the UML or below it,
%giving rise to two characteristic density configurations in the single species model:
%depending on the nutrient and light conditions
%we observe 
%%three characteristic density profiles with 
%either deep or upper phytoplankton maxima, or (in a small parameter regime) an intermediate profile with an essential part of the biomass above and below the thermocline. 
%The third possible phytoplankton profile occur if the parameter changes outside the bistability range. Then we observe a phytoplankton profile with an essential part of the biomass above and below the thermocline. 
%In a range of parameters, both these distributions are stable and their occurrence depends on the initial nutrient distribution.
%This phenomenon, which we refer as the mobility of the production layer, leads to bistability of phytoplankton profiles. 
%
%The inclusion of a UML in the single species model allows for the existence of two alternative density configurations with either deep or upper phytoplankton maxima. 

Inclusion of a UML in the single species model allows for the existence of two alternative density
configurations with either a deep or an upper phytoplankton maximum.
Note that in a small parameter range %close to the critical point 
we observe a third intermediate profile with an essential part of the biomass above and below the thermocline. 
Moreover, there is a range of parameters for which the system is bistable and the 
appearence of either deep or surface biomass maximum
%presence of either of the two spatial configurations 
is determined by the initial distribution of nutrients. 
Both the consumption of nutrient and the self-shading of light are necessary conditions for this behavior. Since the biomass obstructs the upward nutrient flow, it makes the upper layer unfavorable, provided that a deep maximum of phytoplankton has established. The shading of light is an opposite mechanism, which prevents population growth in the lower layers. The third requirement for bistability is the presence of strong mixing in the upper layer. This decouples the locations of the production layer from that of the bulk biomass and prevents a drift of the population toward a DCM, as it would occur in a system without a UML.
%However, in a few recent investigations either surface or deep maxima were observed under almost the same conditions (Venrick 1993; Holm-Hansen and Hewes 2004).
Similar bistable behaviour has also been suggested to occur in field data.
For example, in a recent study of Antarctic waters (Holm-Hansen and Hewes 2004) it was found that adjacent spots could randomly exhibit surface maxima or DCM, even though temperature, salinity and water density profiles of the water columns were practically identical.

%The location of the production layer depends on initial conditions and the presence and distribution of phytoplankton biomass. As a result under certain condition its position within the UML and below it become stable. This phenomena , which we refer as mobility of the production layer, leads to bistability of phytoplankton profiles. 
%In our model bistability of the phytoplankton profile occurs due to the mobility of the production layer, which, in a range of parameters, can be steadily located either within the UML or below it. 

%Inclusion of a UML in the single species model allows for the existence of three characteristic density profiles. If the nutrient is the most limiting factor we observe deep biomass maxima, whereas strong light limitation leads to an upper maximum of biomass. 
%In a range of parameters, both these distributions are stable and their occurrence depends on the initial nutrient distribution.

The occurrence of bistability may have important ecological consequences and usually goes together with catastrophic shifts.
Imagine that we slowly change a parameter, i.e., the incident light intensity, so that its value passes through the bistability range. Then there will be a point, where the current state suddenly loses its stability, thereby inducing a rapid transition from one phytoplankton profile to another. This is also important for a border between oligotrophic and eutrophic waters, where this transition  might be induced by seasonal changes and leads to a shift of the interface between a deep and a surface biomass. maximum. In this scenario bistability will lead to a lag in the transition from one state to another and back.

%Note, that the transition from a deep to surface biomass maximum  and back can occur smoothly if the parameter changes outside the bistability range. Then we observe the third spatial configuration with an essential part of the biomass above and below the thermocline. 

Outside the bistability range the system possesses only one attractor. However, the transition from an unstable to a stable solution may take a long time (especially close to the bistability range) and includes two stages. During the first stage, the nutrient concentration slowly evolves to a level which allows for the formation of a stable biomass profile. This stage occurs on the slow diffusive time scale and may typically last 5-10 years in deep waters. 
Once the nutrient values have crossed a critical value, a rapid transition to the stable profile is triggered. This second stage develops on the biological time scale and lasts approximately 10-50 days.

As a consequence, in this, and in the bistable, regime the system is very sensitive to disturbance. If, due to some perturbation, a large quantity of nutrients is transported into the UML, the system can rapidly switch from a DCM to a UCM and remain in this state for a long time. Similar, a prolonged lack of light can induce a UCM, when the reduced phytoplankton density can not prevent diffusion of nutrients to the surface. 
While outside of the bistability range the reverse transition in principle is possible without further external perturbation, such a flip back to the original state will usually take a long time. 
Thus, the effective size of the bistability regime, i.e. the parameter regime in which two dynamic states are preserved for long times and disturbance can induce long-lasting shifts in the system, is much larger than the true bistability range and comprises a large part of the model's parameter space. Since the relaxation time is long, in a seasonal environment such a system might exhibit only an unstable state (most probable an upper chlorophyll maximum). 

The effects of a UML are even more pronounced in a system that includes two competing species which
are differently adapted to light and nutrient limitation. To analyze this situation we presented a graphical approach which helps to understand the mechanisms of competition in spatially extended systems. 
%of the competition outcome based on the equilibrium distribution of each species alone. 
Moreover, we found that in the range of parameters
where the two species can independently form an upper or a deep biomass maximum, the two species model
demonstrates bistability both in the spatial distribution and in the competition outcome.
Compared to the single-species model, the bistability range is considerably enlarged.
Note that the bistability in the competition outcome is induced by the UML, whereas in a homogeneously mixed water column these species either coexist or only one species wins. 

Another remarkable finding is the survival of a sinking phytoplankton population even if the diffusivity in the deep layers cannot prevent  population washout (Shigesada
and Okubo 1981; Speirs and Gurney 2001; Straube and
Pikovsky 2007). 
While the retention of phytoplankton in the upper layer can be provided by many factors such as the regulation of algal buoyancy or the formation of eddies, etc (Raymont, 1980),
our model shows another dynamic mechanism:
in the absence of phytoplankton in the deep layers, the nutrient can diffuse
upward into the UML, where the population
can start to grow as the nutrient concentration
reaches a sufficient level. However if the light limitation
is not strong enough the biomass shifts
into the deeper layers where it cannot outgrow
the sinking and ultimately declines being limited
by light. Then the nutrient again can diffuse upwards and the cycle repeats.
Thus we identify a novel mechanism of how a population can overcome the drift paradox, as the persistence threshold disappears in the presence of a UML.

%Other mechanisms for the retention of phytoplankton in the upper layer are e.g., regulation of algal buoyancy, formation of eddies, etc (Raymont, 1980).

The mechanisms ensuring persistence ultimately are related to the fact that locally, within the well-mixed upper layer, diffusivity is much larger than the persistence threshold.
Thus obviously, a stationary distribution of a strongly light limited species within the UML will not be affected by low values of diffusivity in deeper layers. However more astonishingly, that the persistence threshold is absent 
%the absence of a persistence threshold prevails also 
for the oscillatory solutions for which the bulk of the biomass remains located in the weakly mixed lower layers for most of the time in the cycle. Again, this effect is related to the mobility of the production layer, which for an oscillatory DCM undergoes cycles in depth. 
These vertical cycles of the production layer can act as an ``oscillatory pump'',  so that 
during some, possibly very small, phase of the cycle the production layer is located within the UML. These short cyclic visits of the UML are sufficient to refuel biomass and to prevent extinction of the population.
Thereby, however, the biomass can be temporarily reduced to very small values 
so that different model outcomes might be expected by including an Allee effect.

Finally, 
%oscillations of the biomass lead to another interesting competitive exclusion effect, where a species can be outcompeted in dependence of its dynamical state.
we have identified an interesting competitive exclusion effect, where a species can be outcompeted in dependence of its dynamical state. 
This occurs for low values of deep diffusivity,
when the DCM of the $N$-species becomes oscillatory (Huisman et al. 2006) and a further reduction of diffusivity favors to the domination of the strongly light limited $I$-species, which still occupies the UML.  At the first glance this result is counterintuitive, because usually a decrease of nutrient transport results in an increased depth of the biomass maximum
%{\bf , an effect which should make the $N$-species, which has lower light requirements, even more robust against a competitor in the UML} 
and here we suddenly obtain a UCM solution instead of a DCM. 
The replacement of the oscillatory solutions occurs because the $I$-species forms a non-oscillatory UCM  
%Now suppose that another species is present in the UML.
and shades light, inhibiting the rapid outburst of the biomass of the $N$-species in the lower layers. As a consequence the (non-oscillatory) UCM configuration in the UML is able to replace the oscillatory deep maximum. 
As some climate models predict higher water stratification with an increase of temperature %by global warming 
(Bopp et al. 2001; Sarmiento et al. 2004),
our findings may help to predict and understand future changes in phytoplankton patterns
ongoing with global climate change.

Possible interesting further extensions of our model would be 
the inclusion of zooplankton or higher trophic levels, the inclusion of an Allee effect or the extension of our study into a three-dimensional system.

\section{Acknowledgments}
We thank Thilo Gross and Wolfgang Ebenh\"{o}h for advise and useful discussions. This study was
supported by German DFG (SFB 555) and German VW-Stiftung.

\appendix

\section{Analytical derivations}

\subsection{Total equilibrium phytoplankton biomass in the single-species model}

\begin{figure}[t]
\begin{center}
\includegraphics[width=8cm]{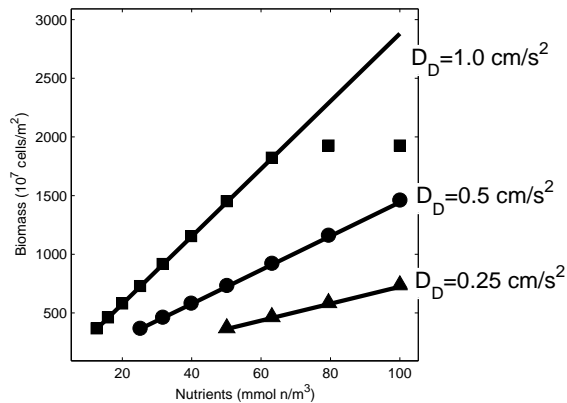}
\end{center}
\caption{Total phytoplankton biomass $W$ as a function of the bottom nutrient concentration $N_B$ for different values of the deep turbulent diffusivity $D_D$. Comparison of the results from \eqref{eq:biomUML} (solid lines) with numerical simulations (symbols).
The analytic estimation is in excellent agreement with the numerical results with exception of the region of high diffusivity $D_D$ and nutrient concentration $N_B$, in which the growth of phytoplankton biomass is limited by light and does not depend on the nutrient concentration.
Thus, the concentration of nutrients in the UML will be larger than $N^*$ and the estimation \eqref{eq:nut_flow_estim} for the nutrient gradient fails.
}\label{fig:biom}
\end{figure}

At equilibrium the left hand sides of Eqs.~\eqref{eq:1sppl} and \eqref{eq:1spntr} equal zero 
\begin{eqnarray}
\mu(N, I) P - m P  - v \mfrac{\partial P}{\partial z}  +
\mfrac{\partial }{\partial z} \left[ D(z) \mfrac{\partial P}{\partial z}\right] &=& 0 
\label{eq:pl_st}\\
\nonumber \\
 -\alpha \mu(N, I) P + \varepsilon \alpha  m P  +
\mfrac{\partial }{\partial z} \left[ D(z) \mfrac{\partial N}{\partial z}\right] &=& 0
\label{eq:nt_st}
\end{eqnarray}
After integration of these equations with respect to the boundary conditions,
Eqs.~\eqref{eq:bound_pl} and \eqref{eq:bound_ntr}, 
\begin{eqnarray}
\int_0^{Z_B}\mu(N, I) P(z) d z - m \int^{Z_B}_0 P(z) dz   & = & 0 
%\label{eq:pl_int}\\
\nonumber \\
-\alpha \int_0^{Z_B}\mu(N, I) P(z) d z + \varepsilon \alpha  m  \int^{Z_B}_0 P(z) dz & = & \nonumber \\
- D(Z_B) \left. \frac{\partial N(z)}{\partial z} \right|_{z=Z_B}   \ ,
\nonumber
%     \label{eq:nt_int}
\end{eqnarray}
we obtain the total biomass
\begin{equation}
 W = \int^{Z_B}_0 P(z) dz =\mfrac{D_D }{\alpha m (1 - \varepsilon)}
\left.  \frac{\partial N(z)}{\partial z} \right|_{z=Z_B} \ .
 \label{eq:totlabiom}
 \end{equation}
This formula relates the phytoplankton biomass and the nutrient flux
and can be interpreted as a conservation law in the system.\\[1ex]
To describe the nutrient flow we first consider the region, which is free of the phytoplankton biomass. Using 
Eq.~\eqref{eq:nt_st} we obtain
\begin{equation}
 D(z) \,  \frac{\partial N(z)}{\partial z}  = \mbox{const} \ .
\end{equation}
This equation corresponds to the stationary diffusion flow and goes together with a linear gradient of nutrients, below the phytoplankton maximum. % (see Fig.~\ref{fig:steady_pl}B). 
If the total phytoplankton biomass is located within the UML, the nutrient flow can be estimated as 
\begin{equation}
D(z) \,  \frac{\partial N(z)}{\partial z} = D_D \mfrac{N_B - N^*}{Z_B - Z_{mix}}  \ ,
\label{eq:nut_flow_estim}
\end{equation}
where $N^*$ from Eq.~\eqref{Eq:limN} approximates the concentration of the nutrient within the UML.
Substituting this into Eq.~\eqref{eq:totlabiom}, we obtain  the total phytoplankton biomass in the UML 
(see Fig.~\ref{fig:biom}).
\begin{equation}
  W = \mfrac{  D_D } {\alpha  m (1 - \varepsilon)} \mfrac{N_B - N^*}{Z_B - Z_{mix}} \ .
  \label{eq:biomUML}
\end{equation}

\subsection{Border of the stability of a UCM \label{sec:bord}}
A sufficient condition for stability of a UCM is the light limitation of growth below the UML, i.e., the light intensity below the UML should be smaller than the critical light intensity $I^*$, see \eqref{Eq:limN}. Using Eq.~\eqref{eq:1splight}, we obtain
\begin{equation}
I_{in} \exp \left[ -K_{bg} Z_{mix} - k \int_0^{Z_{mix} } P(z) d z \right] < I^* \ .
\label{eq:LimLight}
\end{equation}
If the total phytoplankton biomass is located in the UML, equations~\eqref{eq:biomUML} and \eqref{eq:LimLight} give the following criterion for stability of the upper maximum
\begin{equation}
\ln \left( I_{in}/I^* \right) - K_{bg} Z_{mix} <
\mfrac{D_D (N_B - N^*) k}
{ \alpha m (1 - \varepsilon) (Z_B - Z_{mix})} \ .
\label{eq:lbound}
\end{equation}
This line is shown in Figs.~\ref{fig:2sp}A - \ref{fig:2sp}C
as the lower boundary of the bistability range and is in a good agreement with the numerical simulations. 

\subsection{Losses in the UML \label{sec:loss}}
The upper mixed layer is more favorable for sinking phytoplankton species. To show this, consider a water column where the diffusivity does not depend on depth. Its is easy to show that washout from the water column can be interpreted as an additional mortality term.
%%!!
Substituting $P =   \tilde{P}  \exp \left( v z /2D\right)$ into Eq. \eqref{eq:1sppl}, we obtain
$$%%!!
\frac{\partial \tilde{P}}{\partial t} = \mu \tilde{P} - \left( m + \frac{v^2}{4D} \right) \tilde{P}  +
 D \frac{\partial^2 \tilde{P}}{\partial z^2} \ .
$$%%!!
Note that $\partial_t \tilde{P} $ has the same sign as $\partial_t P$, thereby both functions grow and decline simultaneously. Introduce the new mortality 
%%!!
\begin{equation}
m' = m +  \frac{v^2}{4D} \  ,  
\label{eq:adv_mort}
\end{equation}
%%!!
and substitute it in the expression for the limiting resource values \eqref{Eq:limN}.
Since $m' > m$, the new limiting values $I'^*$ and $N'^*$ in the presence of sedimentation should be larger, see \eqref{Eq:limN}. This results in a shift of the zero net growth isoclines towards higher values of resources. However, in a well mixed layer the term $v^2/4D$ vanishes, leading to lower resource requirements. Thus a UML creates more favorable conditions for the sinking phytoplankton biomass.

\linespread{1}
\begin{table*}[p]
\begin{center}
\caption{Parameters values and their meaning}
\begin{tabular}{llll}
\hline
\hline \\[-2ex]
     Symbol & Interpretation  &  Units & Value  \\ [0.5ex]
     \hline\\[-2ex]
     \multicolumn{4}{l}{\textbf{Independent variables}}\\
        $t$ & Time & h & - \\ [0.5ex]
$z$ & Depth & m & - \\ [0.5ex]
\hline\\[-2ex]
\multicolumn{4}{l}{\textbf{Dependent variables}}\\
$P(z, t)$ & Population density & cells m$^{-3}$ &  \\ [0.5ex]
$I(z, t)$ & Light intensity & $\mu$mol photons m$^{-2}$ s$^{-1}$ &  \\[0.5ex]
$N(z, t)$ & Nutrient concentration & mmol nutrient m$^{-3}$ &  \\[0.5ex]
\hline\\[-2ex]
\multicolumn{4}{l}{\textbf{Parameters}}\\
$I_{\rm in}$ & Incident light intensity & $\mu$mol photons m$^{-2}$ s$^{-1}$ & 600 (100 - 600) \\[0.5ex]
$K_{\rm bg}$ & Background turbidity    & m$^{-1}$ & 0.045 \\ [0.5ex]
$k$  & Absorption coefficient of phytoplankton  & m$^2$ cell$^{-1}$ & 6$ \times$10$^{-10}$ \\[0.5ex]
$Z_B$  & Depth of the water column     & m & 300 \\ [0.5ex]
$Z_{\rm mix}$  & Depth of the upper mixed layer  & m & 50  \\[0.5ex]
$w$  & Characteristic width of the thermocline  & m & 1  \\[0.5ex]
$D_D$  & Turbulent diffusivity in the deep layers& cm$^2$ s$^{-1}$ & 0.3 (0.04 - 1) \\ [0.5ex]
$D_U$  & Turbulent diffusivity in the UML& cm$^2$ s$^{-1}$& 50 \\ [0.5ex]
$\mu_{\rm max}$ & Maximum specific growth rate     & h$^{-1}$ & 0.04 \\ [0.5ex]
$H_I$  & Half saturation constant of light  & $\mu$mol photons m$^{-2}$ \ s$^{-1}$ & 20; 98 \\[0.5ex]
   & limited growth for  $N$- and $I$-species &  &  \\ [0.5ex]
$H_N$  & Half saturation constant of nutrient  & mmol nutrient m$^{-3}$ &  0.0425; 0.015\\[0.5ex]
  & limited growth for  $N$- and $I$-species  &  &  \\ [0.5ex]
$m$  & Specific loss rate        & h$^{-1}$ & 0.01 \\[0.5ex]
$\alpha$ & Nutrient content of phytoplankton & mmol nutrient cell$^{-1}$ & 1 $\times$10$^{-9}$ \\[0.5ex]
$\varepsilon$ & Nutrient recycling coefficient & dimensionless & 0.5 \\ [0.5ex]
$v$  & Sinking velocity      & m h$^{-1}$ & 0.042 \\ [0.5ex]
$N_B$  & Nutrient concentration at $Z_B$ & mmol nutrient m$^{-3}$ & 5-100 \\[0.5ex]
     \hline
\end{tabular}
\label{tab:1}
\end{center}
\end{table*}

\begin{table}[p]
\begin{center}
\caption{Acronyms}
\begin{tabular}{ll}
\hline
\hline \\[-2ex]
  Symbol & Interpretation \\ [0.5ex]
  \hline\\ [-2ex]
  DCM & Deep chlorophyll maximum\\ [0.5ex]
 ODCM & Oscillatory or chaotic\\ [0.5ex]
     & deep chlorophyll maximum\\ [0.5ex]
 UCM & Upper chlorophyll maximum\\ [0.5ex]
 UML & Upper mixed layer\\ [0.5ex]
SSC & System state curve\\ [0.5ex]
    \hline \\
\end{tabular}
\label{tab:2}
\end{center}
\end{table}


\begin{thebibliography}{00}

\bibitem{abbott_mixing_1984}
Abbott, M.R., K.L. Denman, T.M. Powell, P.J. Richerson, R.C. Richards, and C.R. Goldman. 1984.
Mixing and the dynamics of the deep chlorophyll maximum in Lake Tahoe.
  Limnol. Oceanogr. 29:862-878.

\bibitem{anderson_subsurface_1969}
Anderson, G.C. 1969.
Subsurface chlorophyll maximum in the northeast Pacific Ocean.
  Limnol. Oceanogr. 14:386-391.

\bibitem{Aristegui03}
Aristegui, J., E.D. Barton, M.F. Montero, M. Garcia-Munoz, and J. Escanez. 2003.
Organic carbon distribution and water column respiration in the {NW}
 Africa-Canaries Coastal Transition Zone.
  Aquatic Microbial Ecology 33:289-301.

\bibitem{Beckmann07}
Beckmann, A. and I. Hense. 2007.
Beneath the surface: Characteristics of oceanic ecosystems under weak mixing conditions - A theoretical investigation.
  Progress In Oceanography 75:771-796.

\bibitem{birch_bounding_2007}
Birch, D.A., Y.-K. Tsang, and W.R. Young. 2007.
Bounding biomass in the Fisher equation.
  Physical Review E 75:066304-14.

\bibitem{bopp_potential_2001}
Bopp, L. et al.
%, P. Monfray, O. Aumont, J.L. Dufresne, H. Le Treut, G. Madec, L. Terray, and J.C. Orr.
2001.
Potential impact of climate change on marine export production.
  Global Biogeochemical Cycles 15:81-100.

\bibitem{Cantrell01}
Cantrell, R.S. and C. Cosner. 2001.
Spatial heterogeneity and critical patch size: area effects via diffusion in closed environments.
J. theor. Biol. 209:161-171.

\bibitem{cullen_deep_1982}
Cullen, J.J. 1982.
The deep chlorophyll maximum: comparing vertical profiles of chlorophyll a.
Can. J. Fish. Aquat. Sci. 39:791-803.

\bibitem{Deuser87}
Deuser, W.G. 1987.
 Variability of hydrography and particle flux: Transient and
  long-term relationships. Mitt Geol-Palaeont Inst Univ Hamburg 62:179-193.

\bibitem{Diehl02}
Diehl, S. 2002.
Phytoplankton, light, and nutrients in a gradient
of mixing depths: theory.
 Ecology 83:386-98.

\bibitem{Finnigan02}
Finnigan, T.D., D.S. Luther, and R. Lukas. 2002.
 Observations of enhanced diapycnal mixing near the Hawaiian ridge.
Journal of Physical Oceanography 32:2988-3002.

\bibitem{Hershey1993}
Hershey, A.E., J. Pastor, B. J. Peterson, and G. W. Kling. 1993.
Stable isotopes resolve the drift paradox for Baetis mayflies in an
arctic river.
Ecology, 74:2315¿2325.

\bibitem{Hodges03}
 Hodges, B.A., and D.L. Rudnick. 2004.
 Simple models of steady deep maxima in chlorophyll and biomass.
 Deep Sea Research I 51:999-1015.

\bibitem{Holm-Hansen04}
Holm-Hansen, O., and C.D. Hewes. 2004.
 Deep chlorophyll-a maxima ({DCM}s) in Antarctic waters I.
  Relationships between DCMs and the physical, chemical, and optical
  conditions in the upper water column.
 Polar Biology 27:699-710.

\bibitem{Holmes94}
Holmes, E.E., M. A. Lewis, J. E. Banks, and R. R. Veit. 1994.
Partial Differential Equations in Ecology: Spatial Interactions and Population Dynamics.
Ecology  75:17-29.

\bibitem{Huisman95}
Huisman, J., and F.J. Weissing. 1995.
Competition for nutrients and light in a mixed water column: a
theoretical analysis.
 The American Naturalist 146:536-564.

\bibitem{Huisman99}
Huisman, J., P. van Oostveen, and F.J. Weissing. 1999.
 Species dynamics in phytoplankton blooms: incomplete
mixing and competition for light.
 The American Naturalist 154:46-48.

\bibitem{Huisman99b}
Huisman, J., P. van Oostveen, and F.J. Weissing. 1999b.
 Critical depth and critical turbulence: two different
mechanisms fordevelopment of phytoplankton blooms.
Limnol. Oceanogr. 44:1781-1787.

\bibitem{Huisman02}
Huisman, J., M. Arrayas, U. Ebert, and B. Sommeijer. 2002.
How do sinking phytoplankton species manage to persist?
The American Naturalist 159:245-254.

\bibitem{Huisman06}
Huisman, J., N.N. Pham Thi, D.M. Karl, and B. Sommeijer. 2006.
 Reduced mixing generates oscillations and chaos in the oceanic deep
  chlorophyll maximum.
 Nature 439:322-325.


\bibitem{Blasius02}
Huppert, A., B. Blasius, and L. Stone. 2002.
A model of phytoplankton blooms.
 The American Naturalist 159:156-171.

\bibitem{Blasius05}
Huppert, A., B. Blasius, R. Olinkya, and L. Stone. 2005.
 A model for seasonal phytoplankton blooms.
 Journal of Theoretical Biology 236:276-290.

\bibitem{jamart_theoretical_1977}
Jamart B.M., D.F. Winter, K. Banse, G.C. Anderson, and R.K. Lam. 1977.
 A theoretical study of phytoplankton growth and nutrient distribution
 in the Pacific Ocean off the northwestern U.S. coast.
 Deep Sea Research 24:753-773.

\bibitem{Jones94}
Jones, C. G., J. H. Lawton, and M. Shachak. 1994. 
Organisms as ecosystem engineers. 
Oikos 69: 373-386.

\bibitem{kierstead_size_1953}
Kierstead, H. and L. B. Slobodkin. 1953.
 The size of water masses containing plankton bloom.
 Mar. Res. 12:141–147.

\bibitem{kirk_light_1994}
 Kirk J.T.O. 1994.
 Light and Photosynthesis in Aquatic Ecosystems.
 Cambridge University Press.

\bibitem{Klausmeier01}
 Klausmeier, C.A., and E. Litchman. 2001.
 Algal games: the vertical distribution of phytoplankton in
  poorly mixed water columns.
 Limnol. Oceanogr. 46:1998-2007.

\bibitem{NatureBoyd2000}
Law, C.S. et al.
%, W.P. Boyd, A.J. Watson, M. Gall, K. Downing, R. Frew, S. Rintoul, S. Pickmere, and R. Pridmore.
2000.
 A mesoscale phytoplankton bloom in the polar Southern Ocean
  stimulated by iron fertilization.
 Nature 407:695-701.

\bibitem{Lewis86}
 Lewis, M.R., W.G. Harrison, N.S. Oakey, D. Hebert, and T. Platt. 1986.
 Vertical nitrate fluxes in the oligotrophic ocean.
 Science 234:870-873.

\bibitem{Matondkar05}
 Matondkar, P.S.G., K.K.C. Nair, and Z.A. Ansari. 2005.
 Biological characteristics of Central Indian Basin waters
  during the southern summer.
 Marine Georesources and Geotechnology 23:299-314.

\bibitem{Neuhauser01}
 Neuhauser, C. 2001.
 Mathematical challenges in spatial ecology.
 Notices of the American Mathematical Society 48:1304-1314.

\bibitem{Okubo01}
Okubo, A. and S.A. Levin 2001. Diffusion and Ecological Problems.
 Springer.

\bibitem{ThiHui05}
 Pham Thi, N.N., J. Huisman, and B.P. Sommeijer. 2005.
 Simulation of three-dimensional phytoplankton dynamics: competition
  in light-limited environments.
 Journal of Computational and Applied Mathematics 174:83-96.

\bibitem{radach_vertical_1975}
Radach, G., and E. Maier-Reimer. 1975.
The vertical structure of phytoplankton growth dynamics; a mathematical model.
Mem. Soc. Roy. Sci. de Liege, 6e serie: 113-146

\bibitem{Raymont}
Raymont, J.E.G. 1980. 
Plankton and productivity in the oceans. I. Phytoplankton.
Pergamon Press, Oxford.

\bibitem{riley_quantitative_1949}
Riley, G. A. and H. Stommel and D. F. Bumpus. 1949.
Quantitative ecology of the plankton of the Western North Atlantic.
Bull Bingham Oceanogr. Coll 12:1-169.

\bibitem{Ryabov2008}
Ryabov, A. B. and B. Blasius. 2008. 
Population growth and persistence in a heterogeneous environment: the role of diffusion and advection.
Math. Model. Nat. Phenom. 3:42-86.

\bibitem{Saggio2001}
Saggio, A. and J. Imberger. 2001. 
PMixing and turbulent fluxes in the metalimnion of a stratified lake.
Limnol. Oceanogr. 46:392-409.

\bibitem{sarmiento_response_2004}
Sarmiento, J. L. et al. 2004.
%and R. Slater and R. Barber and L. Bopp and S. C. Doney and A. C. Hirst and J. Kleypas and R. Matear and U. Mikolajewicz and P. Monfray.
Response of ocean ecosystems to climate warming.
Global Biogeochem. Cycles 18:GB3003, doi:10.1029/2003GB002134.

\bibitem{Shigesada81}
Shigesada, N., and A. Okubo. 1981.
Effects of competition and shading in planktonic communities.
Journal of Mathematical Biology 12:311-326.

\bibitem{skellam_random_1951}
Skellam, J.G. 1951.
Random dispersal in theoretical populations.
Biometrika 38:196-218.

\bibitem{Speirs01}
Speirs, D.C., and W.S.C. Gurney. 2001.
Population persistence. in rivers and estuaries.
Ecology 82:1219-1237.

\bibitem{Smyth01}
Smyth, W.D., J.N. Moum, and D.R. Caldwell. 2001.
 The efficiency of mixing in turbulent patches: inferences from direct
  simulations and microstructure observations.
 Journal of Physical Oceanography 31:1969-1992.

\bibitem{Sverdrup53}
Sverdrup, H. U. 1953.
On conditions for the vernal blooming of phytoplankton.
J. Conseil. 18:287-295.

\bibitem{steele_vertical_1960}
Steele, J.H. and C.S. Yentsch 1960.
The vertical distribution of chlorophyll.
J. Mar. Biol. Assoc. UK 39:217-226.

\bibitem{Straube07}
Straube, A.V., and A. Pikovsky. 2007.
Mixing-induced global modes in open active flow.
Phys. Rev. Lett. 99:184503.

\bibitem{Tilman80}
Tilman, D. 1980.
Resources: a graphical-mechanistic approach to competition and predation.
The American Naturalist 116:362-93.

\bibitem{Tilman82}
Tilman, D. 1982. Resource competition and community structure. New York: Princeton Univ. Press.

\bibitem{Tilman97}
Tilman, D. and P. M. Kareiva 1997.
Spatial Ecology: The role of space in population dynamics and interspecific interactions.
New York: Princeton Univ. Press.

\bibitem{Tittel03}
Tittel, J., V. Bissinger, B. Zippel, U. Gaedke, E. Bell, A. Lorke, and N. Kamjunke. 2003.
 Mixotrophs combine resource use to outcompete
specialists: Implications for aquatic food webs
 PNAS 100:12776-12781.

\bibitem{turpin_d.h._physiological_1988}
Turpin, D.H. 1988.
Physiological mechanisms in phytoplankton resource competition. In: Sandgren, C. D. (ed.)
Growth and reproductive strategies of freshwater phytoplankton.
Cambridge Univ. Press, Cambridge:316-368.

\bibitem{varela_modelling_1992}
Varela, R.A., A. Cruzado, J. Tintore, and E.G. Ladona. 1992.
Modelling the deep-chlorophyll maximum: A coupled physical-biological approach.
Journal of Marine Research. 50:441-463.

\bibitem{Venrick93}
Venrick, E.L. 1993.
Phytoplankton seasonatity in the central North Pacific: The
 endless summer reconsidered.
 Limnol. Oceanogr.. 38:1135-1149.

\bibitem{Weston05}
Weston, K., L. Fernand, D.K. Mills, R. Delahunty, and J. Brown. 2005.
 Primary production in the deep chlorophyll maximum of the central
  North Sea.
 Journal of Plankton Research 27:909-922.

\bibitem{Yoshiyama2002}
Yoshiyama, K., and H. Nakajima. 2002.
 Catastrophic transition in vertical distributions of phytoplankton:
alternative equilibria in a water column.
 Journal of Theoretical Biology 216:397-408.

\bibitem{Yoshiyama2006}
Yoshiyama, K., and H. Nakajima. 2006.
 Catastrophic shifts in vertical distributions
 of phytoplankton. The existence of a bifurcation set.
 Journal of Mathematical Biology 52:235-276.


\end{thebibliography}
\end{document}